\documentclass[12pt]{article}
\usepackage{epsfig}
\usepackage{amsmath,amssymb,amsthm,amscd}
\def\IP{{\mathbb P}}
\def\re{{\rm e}}
\def\ZZ{{\mathbb Z}}
\def\IR{{\mathbb R}}
\def\IC{{\mathbb C}}
\def\bT{{\mathbb T}}
\def\IQ{{\mathbb Q}}
\def\bC{{\mathbb C}}

\def\2{{1\over2}}
\let\<=\langle \let\>=\rangle
\def\new#1\endnew{{\bf #1}}
\def\ifundefined#1{\expandafter\ifx\csname#1\endcsname\relax}
\ifundefined{draftmode}\else    \input draftmode        \fi
\let\Msize=\footnotesize             
\def\BM{\Msize\begin{matrix}}           \def\EM{\end{matrix}}
\def\MN M:#1 #2 N:#3 #4 {{(#1_{#2},#3_{#4})}}
\def\MNH M:#1 #2 N:#3 #4 H:#5,#6 [#7]{{(#1_{#2},#3_{#4})^{#5,#6}_{#7}}}

\newcommand{\ds}{\displaystyle}

\def\dd{\mathrm{d}}

\newcommand{\CS}{{\cal S}}
\newcommand{\CP}{{\cal P}}
\newcommand{\ri}{{\rm i}}

\newcommand{\be}{\begin{equation}}
\newcommand{\ee}{\end{equation}}
\newcommand{\ba}{\begin{aligned}}
\newcommand{\ea}{\end{aligned}}
\newcommand{\ben}{\begin{eqnarray}\displaystyle}
\newcommand{\een}{\end{eqnarray}}

\newcommand{\half}{{1\over 2}}
\newcommand{\sectiono}[1]{\section{#1}\setcounter{equation}{0}}

\begin{document}

\begin{titlepage}
{}~
\hfill\vbox{
\hbox{CERN-PH-TH/2005-258}
\hbox{MAD-TH-05-9}
}\break

\vspace{1cm}

\centerline{\Large \bf
Counting BPS states on the Enriques Calabi-Yau}

\vspace*{5.0ex}

\centerline{ \large \rm Albrecht Klemm$^a$ and
Marcos Mari\~no$^b$\footnote{Also at Departamento de Matem\'atica, IST, Lisboa, Portugal} }

\vspace*{4.0ex}

\centerline{ \rm ~$^a$UW-Madison Physics Department,
1150 University Avenue }
\centerline{\rm Madison, WI 53706-1390, USA}
\centerline{\tt aklemm@physics.wisc.edu}

\vspace*{1.8ex}

\centerline{ \rm ~$^b$Department of Physics, Theory Division, CERN}

\centerline{ \rm CH-1211 Geneva, Switzerland}

\centerline{\tt
marcos@mail.cern.ch}

\vspace*{6ex}

\centerline{\bf Abstract}
\medskip
We study topological string amplitudes for the FHSV model using various techniques. This model
has a type II realization involving a Calabi-Yau threefold with Enriques fibres, which we call the
Enriques Calabi-Yau.
By applying heterotic/type IIA duality, we compute the topological amplitudes in the fibre
to all genera. It turns out that there are two different ways to do the computation that lead to
topological couplings with different BPS content. One of them gives the standard D0-D2 counting
amplitudes, and from the other one we obtain information about
bound states of D0-D4-D2 branes on the Enriques fibre. We also study the model using mirror symmetry and
the holomorphic anomaly equations. We verify in this way the heterotic results for the D0-D2 generating functional
for low genera and find closed expressions for the topological amplitudes on the total
space in terms of modular forms, and up to genus three. This model turns out to be
much simpler than the generic B-model and might be exactly solvable.

\end{titlepage}
\vfill
\eject


\newpage

\baselineskip=16pt

\tableofcontents

\sectiono{Introduction}
\label{sec:introduction}
The solution of topological string theory on Calabi-Yau (CY) manifolds
is an important problem with applications both in string theory and in
enumerative geometry. Impressive tools have been developed in an extensive
effort over the last twelve years, most notably mirror symmetry, localization,
deformation, large $N$-dualities, cohomological calculations in D2-D0 brane
moduli spaces, and heterotic/type II duality. Nevertheless, a complete
solution for the topological string amplitudes on compact CY manifolds is
presently out of reach\footnote{In contrast, on non-compact toric Calabi-Yau manifolds
the problem is completely solved by localization, and much more efficiently by using the
topological vertex \cite{akmv}, which relates the genus expansion to a $1/N$
expansion of Chern-Simons theory.}.
Applicable to the compact case are B-model calculations based on
mirror symmetry and the holomorphic anomaly equation of \cite{bcov,kkv}, and A-model
calculations based on deformation arguments and relative Gromov-Witten
invariants, which are calculated by localization \cite{Gathmann,mppaper}.
Both methods calculate the amplitudes genus by genus. The former provides
contributions in all degrees at once, but only  up to a holomorphic function
(the so-called holomorphic ambiguity), whose determination requires
further finite amount of data, which in practice are provided in a rather unsystematic
case by case analysis. The A-model calculation proceeds degree by degree
and the combinatorial complexity is in general prohibitive. Many 
(potentially all) compact CY manifolds are connected by transitions 
through complex degenerations. The behaviour of the topological string 
amplitudes at these transitions is relatively well understood. In 
view of this situation it is important to identify {\sl the} compact CY 
manifold where the topological string is most tractable.

There is a compact example where topological string
theory is exactly solvable, namely K3$\times \bT^2$.
The topological string amplitudes are all zero for genus $g\ge 2$, at $g=0$ one has
just the classical piece of the prepotential, and for $g=1$ one just has
the elliptic $\eta$ function typical of the two-torus \cite{bcov1}. Hence this
example is too simple, and this is due to the extended ${\cal N}=4$ supersymmetry
of the corresponding type II theory, related in turn to the $SU(2)$ holonomy.
${\cal N}=2$ supersymmetry and the generic $SU(3)$ holonomy can be obtained by
fibering K3 over $\IP^1$. In these examples one can use
heterotic/type II duality or special properties of the Hilbert scheme
of complex surfaces to write down explicitly all genus topological amplitudes
for the classes in the K3 fiber \cite{mm,kkrs}. However the
decisive step in going from the surface to the threefold, i.e. the inclusion
of the base and mixed classes, is hard. Results up to $g=2$ have been obtained in \cite{kkrs}.

This motivates to consider the problem on a special CY with intermediate holonomy
$SU(2)\times \ZZ_2$ constructed in \cite{Borcea,Voisin,fhsv}, as an
orbifold w.r.t. a free $\ZZ_2$ involution of K3$\times \bT^2$.  
The resulting space exhibits a K3 fibration with four fibres of multiplicity two
over the four fixed points of the involution in the base, which are Enriques
surfaces. A good deal of the nontrivial geometry of this CY comes from the geometry of the
Enriques fibers, and we will call it the {\it Enriques CY manifold}. The string vacuum obtained
by compactifying type II theory on the Enriques CY has ${\cal N}=2$ supersymmetry
and is known as the FHSV model. The $\ZZ_2$ lifts instanton zero modes related 
to the $\bT^2$ so that simple instanton effects which cancel in the ${\cal N}=4$ theory 
contribute to the ${\cal N}=2$ effective action. 
The Enriques CY seems to be the simplest CY compactification with 
nontrivial topological string amplitudes. Moreover it has a dual 
description as an asymmetric orbifold of the heterotic string 
\cite{fhsv}. Various aspects of the FHSV model have been studied in 
the past, see for example \cite{aspinwall,bdm}. In particular, the genus 
one topological string amplitude of the FHSV model was determined by 
Harvey and Moore in \cite{hmborcherds}.

This paper makes a first step to determine the topological string amplitudes of the FHSV model
using heterotic/type II duality, B-model techniques and the cohomology of D2-D0 brane moduli
spaces. Although we haven't been able to solve the model in full,
we will present various results which show that indeed the model has some simplifying
features that might lead to a complete solution.  The simplicity of the model is also
apparent from a mathematical point of view, and it turns out that the techniques
developed in \cite{Gathmann,mppaper} lead to simple recursive formulae for the
Gromov-Witten invariants of the Enriques CY at low genera \cite{mp}.

A  heterotic one-loop calculation is used to determine the $F_g$ couplings in the fibre direction,
using the techniques developed in \cite{hm,agnt,mm}. It turns out that this calculation can be
made in two different ways, which we call the {\it geometric reduction} and the {\it Borcherds-Harvey-Moore (BHM)
reduction}. The resulting expressions are appropriate for different regions in moduli space,
and they turn out to have a different enumerative meaning. The result obtained in the geometric
reduction corresponds to the large radius limit, reproduces the geometric expectations one has
for a generating functional of Gromov-Witten invariants, and as explained in \cite{gv} counts
D0-D2 bound states. We suggest that the result obtained in the BHM reduction is related to a counting
of D0-D2-D4 bound states in a different region of moduli space. The result of
\cite{hmborcherds} for the genus one amplitude was in fact implicitly obtained in the
BHM reduction.

We then study the model by using mirror symmetry and the holomorphic anomaly equations.
To do that, we first find an algebraic realization of the CY manifold involved in the FHSV model, and
we find its mirror by using standard techniques. This leads to a model which is still
very difficult to solve due to the presence of ten deformation parameters. To avoid this
problem, we find a reduced model with only two fibre parameters which is obtained by blowing down
the $E_8$ part of the homology of the original type A model. This model turns out to be very tractable,
and all relevant quantities can be expressed in closed form in terms of modular forms. We present
explicit formulae for the topological string amplitudes in the fiber up to genus $3$ which
agree with the predictions of heterotic/type II duality. The holomorphic anomaly equations turn out
to be extremely simple due to various exceptional properties of the model (like the absence of worldsheet instanton
corrections for the prepotential, already pointed out in \cite{fhsv}). Although we do not have enough information
to fix the holomorphic ambiguity in the base (except at genus 2, where explicit results have been obtained in \cite{mp}), we
make a natural conjecture that leads to consistent results in genus three and four and might hold in general.

The organization of this paper is as follows. In section 2 we review the heterotic computation of topological
string amplitudes in K3$\times \bT^2$ compactifications. In section 3 we present various results on the FHSV model,
both in its type IIA and its heterotic incarnations. In section 4 we compute the topological string amplitudes in the
heterotic theory, in both the geometric and the BHM reductions. In section 5 we present an interpretation of the
results in the geometric reduction in terms of BPS invariants associated to D2-D0 bound states, following \cite{gv,kkv}.
In section 6 we study the mirror B-model for a ``reduced'' version of the theory, and we compare the results
with those obtained in the heterotic computation. Appendix A collects some useful formulae for modular forms. Appendix B
summarizes some results about the lattice reduction technique which is used in the heterotic computation.

\sectiono{Heterotic/type II duality and $F_g$ couplings}
\label{sec:heterotictypeII}

\subsection{The $F_g$ couplings in heterotic string theory}

The duality between heterotic compactifications on K3$\times \bT^2$ and type II theory on Calabi-Yau's which are
K3 fibrations \cite{kv} has been a source of very rich information in string theory (see for example \cite{lust} for a review of
results). One of the most interesting applications of this duality is the computation of the topological string amplitudes
$F_g$. These $F_g$ couplings are F-terms for compactifications of type II theory on Calabi-Yau manifolds
\cite{bcov,atop}, and they give terms in the four-dimensional effective action of the form
\be
\int F_g(t, \bar t) T^{2g-2} R^2 + \cdots
\ee
where $T$ is the graviphoton field strength and $R$ is the Riemann curvature. It turns out that, on the
heterotic side, all these couplings appear at one-loop \cite{agnt} and can be computed in closed form
\cite{mm}. In this section we will briefly review the computation of the $F_g$ amplitudes by using heterotic/type II duality.
One drawback of this duality is that it only enables us to compute this amplitude in the limit of infinite volume
for the basis of the K3 fibration. This is due to the fact that, under heterotic/type IIA duality, the heterotic dilaton $S$
is identified with the complexified area of the base $\IP^1$ of
the fibration,
\be
{\rm vol}_{\IC}(\IP^1)={4 \pi S}={4 \pi\over g_{\rm het}^2}.
\label{eq:limit}
\ee
Therefore the perturbative regime of the heterotic string corresponds to the limit
${\rm vol}_{\IC}(\IP^1) \rightarrow \infty$. On the other hand,
the duality gives closed, elegant expressions for all the $F_g$ amplitudes restricted to fiber classes
in terms of modular forms. These classes correspond to the Picard lattice of the K3 fibre, which will be
denoted by ${\rm Pic}({\rm K3})$ (more precisely, one has to consider the monodromy-invariant part of the Picard lattice).

Before stating the main results, we introduce some notation on Narain lattices and Siegel-Narain theta functions.
Given a lattice $\Gamma$ of signature $(b_+, b_-)$,
a projection $P$ is an orthogonal decomposition of $\Gamma \otimes \IR$ into subspaces of definite signature:
$$
P: \Gamma \otimes \IR \simeq \IR^{b_+} \perp \IR^{b_-}.
$$
We will denote by $p_{\pm}=P_{\pm }(p)$ the projections onto the two factors.
The Siegel-Narain theta function is defined as
\be
\Theta_{\Gamma}(\tau, \alpha, \beta) =
\sum_{p \in \Gamma}\exp\biggl\{ \pi i \tau(p+\beta/2)^2_+ + \pi i {\overline \tau}
(p+\beta/2)^2_- + \pi i (p + \beta/2,\alpha)\biggr\}.
\ee
When $\alpha=\beta=0$, we will simply write $\Theta_{\Gamma}(\tau)$. As usual, we write $q=\exp(2\pi i \tau)$, and
$\tau_2={\rm Im}\, \tau$.

We will consider compactifications of the heterotic string on K3$\times \bT^2$ and orbifolds thereof. These compactifications lead
to effective theories with ${\cal N}=2$ supersymmetry in four dimensions, and they involve Narain lattices with $b^+=2$, which can be
identified with the two right-moving directions along $\bT^2$. Therefore,
we can identify $\IR^{b_+}\simeq {\bC}$, and we will
represent $p_+ \in \IR^2$ by $p_R \in {\bC}$, so that $p_+^2$ is given by $|p_R|^2$.

The general expression for the $F_g$ couplings in these
compactifications is given by the one-loop integral \cite{agnt} (see \cite{kiritsis} for an excellent
introduction to one-loop corrections in string theory)
\be
\label{Fg}
F_g=\int_{\cal F} {d^2 \tau  \over \tau_2} {1\over |\eta|^4}\sum_{\rm even} {i\over \pi} (-1)^{a + b + ab} \partial_{\tau}
\biggl( {\vartheta[^a_b](\tau) \over \eta(\tau)}\biggr) Z_g^{\rm int}[^a_b].
\ee
In this equation, the integration is over the fundamental domain of the torus,
\be
Z_g^{\rm int}[^a_b]=\langle :\bigl( {\partial X}\bigr)^{2g}: \rangle
\ee
is a correlation function evaluated in the internal conformal field theory, and $X$ is the complex boson corresponding to the right-moving
modes on the $\bT^2$. The evaluation of the correlation function reduces to zero modes \cite{agnt}, and
the final result involves insertions of the right-moving
momentum $p_R$. For this reason, it is convenient to introduce the Narain theta function
with an insertion,
\be
\Theta_{\Gamma}^g(\tau, \alpha, \beta)=
\sum_{p \in \Gamma}  p_R^{2g-2}\exp\biggl\{ \pi i \tau(p+\beta/2)^2_+ + \pi i {\overline \tau}
(p+\beta/2)^2_- + \pi i (p + \beta/2,\alpha)\biggr\}.
\ee
In general the internal CFT will be an orbifold theory and we will have to consider different orbifold blocks, which will be labelled by
$J$. For each of these blocks there is a different Narain lattice $\Gamma_J$ with different $\alpha_J, \beta_J$, and
we will denote
\be
\label{thetaJ}
\Theta^g_J=\Theta^{g}_{\Gamma_J}(\tau, \alpha_J, \beta_J).
\ee
The integral (\ref{Fg}) can now be written as \cite{agnt, ms}
\be
\label{Fgint}
F_g=\int_{\cal F} d^2 \tau  \tau_2^{2g-3} \sum_{J} {\cal I}^g_J,
\ee
where
\be
\label{gintegrand}
{\cal I}^g_J= {{\widehat {\cal P}}_{g} (q)\over Y^{g-1}} {\overline \Theta}^g_J(\tau)f_J(q).\\
\ee
In this equation,
${\widehat {\cal P}}_{g}(q)$ is a one-loop correlation function of the bosonic fields and is given by \cite{lnsw,agnt}
\be
\label{defphatg}
\re^{-\pi \lambda^2 \tau_2} \biggl( { 2\pi  \eta^3 \lambda \over \vartheta_1(\lambda|\tau)}\biggr)^2=
\sum_{g=0}^{\infty} (2 \pi \lambda)^{2g} {\widehat {\cal P}}_{g}(q).
\ee
$f_J(q)$ is a modular form which depends on the
details of the internal CFT. Finally,
the quantity $Y$ in (\ref{gintegrand}) is a moduli-dependent function related to the K\"ahler potential as $K=-\log \, Y$.
We will also define the holomorphic counterpart ${\cal P}_g (q)$ of ${\widehat {\cal P}}_{g}(q)$ by
\be
\label{defpg}
\biggl( { 2\pi  \eta^3 \lambda \over \vartheta_1(\lambda|\tau)}\biggr)^2=
\sum_{g=0}^{\infty} (2 \pi \lambda)^{2g} {\cal P}_{g}(q).
\ee
The quantities ${\cal P}_{g}(q)$ can be explicitly written in terms of generalized Eisenstein series. To do this, one uses the expansion
\be
{ 2\pi \eta^3 z \over \vartheta_1 (z|\tau) } = -\exp \Biggl[
\sum_{k=1}^{\infty} {\zeta(2k) \over k}
E_{2k} (\tau) z^{2k}\Biggr].
\ee
If we now introduce the polynomials $\CS_k$ through:
\be
\label{schur}
\exp\Bigl[ \sum_{n=1}^\infty x_n z^n \Bigr]
= \sum_{n=0}^\infty \CS_n(x_1, \dots, x_n) z^n,
\ee
we can easily check that  $\CP_{g}(q)$ is a quasimodular form of weight $(2g,0)$ given
by
\be
\label{bigpoly}
\CP_{g}(q) = \CS_{g}\biggl( x_k={|B_{2k}|\over k (2k)!} E_{2k}(q)\biggr).
\ee
where $B_{2k}$ are Bernoulli numbers, and $E_{2k}(q)$ is the Eisenstein series introduced in (\ref{geneis}).
We have, for instance,
\be
\label{casesps}
\CP_1(q)={1\over 12}E_2(q), \,\,\,\,\,\, \CP_2(q)={1 \over 1440} (5 E_2^2 + E_4).
\ee
The non-holomorphic modular forms ${\widehat {\cal P}}_g$ are obtained by an equation identical to (\ref{bigpoly}) after changing
$E_2$ by
\be
{\widehat E}_2(\tau) =E_2(\tau)-{3\over \pi \tau_2}.
\ee

The computation of the $F_g$ amplitudes involves, then, the determination of the modular forms $f_J(q)$, and the evaluation of the
integral over the fundamental domain. The first step is easy when there is an orbifold realization. The second step is more involved
and requires the method first introduced in \cite{dkl} in the
context of string threshold corrections. This method was further refined and developed in
\cite{hm,borcherdstwo, mw}. We will refer to the approach presented in \cite{hm,borcherdstwo,mw} to calculate these
integrals as the {\it lattice reduction} technique. This technique, which was used to compute the $F_g$
couplings in \cite{mm}, computes the integral (\ref{Fgint}) iteratively by
``integrating out'' a sublattice of the Narain lattice of signature $(1,1)$, therefore
reducing its rank at every step. In the cases considered in this paper, where one starts with
lattices of signature $\Gamma^{2,2+s}$, it is sufficient to
perform the lattice reduction once by ``integrating out'' a sublattice
\be
\Gamma^{1,1}=\langle z, z'\rangle.
\ee
The generating vector $z$ is called the reduction vector. The details of the lattice
reduction procedure are rather intricate, and we present some of them in Appendix B. There are two general important
properties of the result for $F_g$ which are worth pointing out. The first one is that different choices for the sublattices
$\Gamma^{1,1}$ to be integrated out in the process of lattice reduction lead in general to different results for the
integral, and each of these expressions is valid in a different region of moduli space. The second property is that,
although the result for the integral (\ref{Fgint}) is rather complicated, the holomorphic limit
\be
\bar t \rightarrow \infty, \qquad t \,\, {\rm fixed},
\ee
leads to a rather simple expression for $F_g$. This holomorphic limit is the one needed to
extract the topological information of $F_g$ \cite{bcov}.

We now present some general results on the holomorphic limit of $F_g$, obtained from a
lattice reduction computation of the heterotic integral (\ref{Fgint}). For simplicity, we will restrict
ourselves to the case in which one has a single lattice involved in the
integrand (\ref{Fgint}), and $\alpha=\beta=0$.
In general, the integrand will be a sum over different orbifold blocks and different lattices $\Gamma_J$ with nonvanishing $\alpha, \beta$. The final
answer for $F_g$ in these cases will be given by a sum over the different blocks. The presence of $\alpha, \beta$ leads however to nontrivial modifications
of the result, as it has been already noticed in various papers \cite{ms,hem, mw,mmtwo,lm}. We will consider these modifications when we
analyze the FHSV model. Most of the results we will present have been
obtained in \cite{mm}, although we will consider a slightly more general situation which will be needed
for the FHSV model.

We first introduce some necessary ingredients to write down the answer.
First of all, the norm $|z_+|^2$ of the projected reduction vector depends on the Narain moduli of
the compactification as
\be
\label{zquot}
|z_+|^2 ={\nu \over Y},
\ee
where $Y=\re^{-K}$ is the moduli-dependent quantity introduced in (\ref{gintegrand}), and $\nu$
is a real number related to the norm of $z$. In the STU model considered in \cite{mm}, $\nu=1$, but
as we will see in general it can take other values. We will label an element $p^K$ of the reduced lattice by
a vector $r$ of integer coordinates. The resulting expression for the holomorphic limit of $F_g$ depends on the
moduli through the combination \cite{mm}
\be
\label{oneargument}
\exp\biggl( 2\pi \ri (p^K, \mu/N) + {2\pi\widetilde P_+ (p^K)\over |z_+|}\biggr).
\ee
The different ingredients in this expression are explained in detail in Appendix B. The first term of (\ref{oneargument})
comes from the exponent in the second line of (\ref{nondeg}), and
the second term comes from the argument of the Bessel function in (\ref{nondeg}). One
can easily seen, by using the explicit expressions for the different quantities involved,
that the exponent in (\ref{oneargument}) can be written in the form
\be
\nu^{-{1\over 2}} 2\pi \ri r\cdot y,
\ee
where $y$ is a vector of holomorphic coordinates for the heterotic moduli space
(which is related to $t$, the flat coordinates in the positive K\"ahler cone in the
type II realization, in a simple way). We will see concrete examples of this
in the calculation for the FHSV model. Finally, we define the coefficients $c_g(n)$ through
\be
\label{defcg}
{\cal P}_{g} (q) f(q)=\sum_n c_g(n) q^n.
\ee
The final expression for $F_g$ is:
\be
\label{fgex}
F_g(t)=\nu^{1-g}  \sum_{r>0} c_g(r^2/2)\sum_{\ell=1}^{\infty} \ell^{2g-3} {\rm e}^{\ell \nu^{-{1\over 2} } 2\pi \ri r \cdot y}.
\ee
In this equation, $r^2$ is computed with the norm of the reduced lattice $K$, and the restriction $r>0$ means that we
consider vectors such that ${\rm Im}(r\cdot y)>0$, as well as a finite number of boundary cases \cite{hm,mm}. The sum over $\ell$ in (\ref{fgex}) can be written
as
\be
{\rm Li}_{3-2g}({\rm e}^{ \nu^{-{1\over 2} } 2\pi \ri r \cdot y}),
\ee
where ${\rm Li}_{n}$ is the polylogarithm of index $n$ defined as
\be
{\rm Li}_n (x) =\sum_{k=1}^{\infty} {x^k \over k^n}.
\ee
In the above expression for $F_g$ we are not taking into account constant terms as well as
polynomial terms in $y$ and ${\rm Im}\, y$ which also appear in the heterotic computation \cite{hm,mm}.

\subsection{BPS content of the $F_g$ couplings}

As shown in \cite{hm,hmal}, the couplings $F_g$ are BPS-saturated amplitudes and they can be regarded as
generating functions that count in an appropriate way the BPS states of the ${\cal N}=2$ compactification.
The underlying structure of the couplings was further clarified in the work of Gopakumar and Vafa \cite{gv}, who gave a
precise formula for the BPS content of the $F_g$ in terms of bound states of D0-D2 branes in a type IIA compactification
on a CY threefold $X$. These bound states lead to BPS particles in four dimensions labelled by three quantum numbers. The first quantum number
is the homology class $r\in H_2(X, \ZZ)$ of the Riemann surface wrapped by the D2. The other two quantum numbers are given by the off-shell
spin content $j_L$, $j_R$ with respect to the algebra ${\rm su}(2)_L\times  {\rm su}(2)_R$ of the rotation group $SO(4)$.  Let us denote by
$N_{j^3_L,j^3_R}(r)$ the number of BPS states with these quantum numbers. This number is not invariant under deformations, therefore
\cite{gv} considered the index  $n_g(r)$ defined by
\be
\sum_{j^3_L,j_R^3} (-1)^{2 j_R^3}(2 j_R^3+1) N_{j^3_R,J^3_L}(r)[{\bf j_L}]=
\sum_{g=0}^\infty n_g(r) I_g\ ,
\label{index}
\ee
where $I_g=\left[\left({\bf 1\over 2}\right)_L
+2 ({\bf 0})_L\right]^{\otimes g}$. The integer numbers $n_g(r)$, which characterize the spectrum of D2-D0 bound states
in CY compactifications of type IIA, are called {\it Gopakumar-Vafa (GV) invariants}.

Let us now consider the generating functional of
topological string amplitudes
\be
F(\lambda)=\sum_{g=0}^{\infty} F_g(t) \lambda^{2g-2}.
\ee
According to \cite{gv}, the worldsheet instanton corrections to $F(\lambda)$ can be obtained by a Schwinger one-loop computation involving only the D2-D0 bound states:
\be
 \label{gvloop}
F(\lambda)=\sum_{g=0}^{\infty}   \sum_{r \in H_2(M,\ZZ)}\sum_{m=-\infty}^{\infty}  n_g(r) \int_0^{\infty} {ds \over s} \biggl( 2 \sin {s\over 2}\biggr)^{2g-2}
 \exp\Bigl[ -{s \over \lambda}(r\cdot t + 2\pi \ri m)\Bigr].
 \ee
In this formula, the sum over $m$ is over the number of D0 states bound to the D2s, and we have taken into account that the index $n_g(r)$
is independent of $m$. After a Poisson resummation over $m$ one finds \cite{gv}:
\be
F(\lambda)=\sum_{g=0}^\infty  \sum_{r \in H_2(M,\ZZ)} \sum_{d=1}^{\infty}
n_g(r) {1\over d}
\left(2 \sin {d \lambda \over 2}\right)^{2 g-2}  {\rm e}^{-d r\cdot t},
\label{gova}
\ee
Notice that the sum over $d$ in (\ref{gova}) plays the same role as the sum over $\ell$ in the heterotic computation
(\ref{fgex}). This expression, which takes into account the spectrum of  ``electric" states associated to D2-D0 branes,
is valid in the large radius limit of the CY compactification, since in this region the lightest
states are indeed the D2 and D0 branes and their bound states, while the D4 and D6 ``magnetic" states are heavy.
Equation (\ref{gova}) leads to strong structural predictions for the topological string amplitudes $F_g$ when
written in terms of GV invariants. Up to genus $4$, one finds (for the instanton part)
\be
\label{gvexpansion}
\ba
F_0=&\sum_{r \in H_2(M,\ZZ)} n_0(r) {\rm Li}_3(\re^{-r\cdot t}),\\
F_1=&\sum_{r \in H_2(M,\ZZ)} \biggl( {n_0(r) \over 12} + n_1(r) \biggr) {\rm Li}_1(\re^{-r\cdot t}),\\
F_2=&\sum_{r \in H_2(M,\ZZ)} \biggl( {n_0(r) \over 240} + n_2(r) \biggr) {\rm Li}_{-1} (\re^{-r\cdot t}),\\
 F_3=&\sum_{r \in H_2(M,\ZZ)}\biggl( {n_0(r) \over 6048}- {n_2(r) \over 12}+
n_3(r) \biggr){\rm Li}_{-3}(\re^{-r\cdot t}),\\
F_4=&
\sum_{r \in H_2(M,\ZZ)}\biggl({n_0(r) \over 172800}+{n_2(r) \over 360}-{n_3(r) \over 6}+n_4(r) \biggr){\rm Li}_{-5}(\re^{-r\cdot t}).
\ea
\ee

 In the simple case where $\nu=1$, the heterotic result (\ref{fgex}) leads to a simple generating function for the GV invariants. To see
 this, notice that, if we write
 \be
 F(\lambda)=\sum_{g=0}^\infty  \sum_{r \in H_2(M,\ZZ)} \widehat N_{g}(r)
{\rm Li}_{3-2g}( {\rm e}^{-r\cdot t}) \lambda^{2g-2},
\label{gvinter}
\ee
then from (\ref{gova}) one has the following relation for fixed $r$
\be
\sum_{g=0}^{\infty} n_g(r)
\left(2 \sin { \lambda \over 2}\right)^{2 g-2} =\sum_{g=0}^{\infty}\widehat N_{g}(r) \lambda^{2g-2},
\ee
Under heterotic/type II duality, and with an appropriate choice of lattice reduction,
the reduced lattice $K$ that appears in the heterotic computation becomes the Picard lattice of the
K3 fiber ${\rm Pic}({\rm K3})$, and the vectors $r$ label homology classes in this lattice.
According to (\ref{fgex}), one has that $\widehat N_g(r)=c_g(r^2/2)$, where the $c_g(n)$ are
defined in (\ref{defcg}). If we now use (\ref{defpg}) and
the product  representation of $\vartheta_1(\nu|\tau)$ given in (\ref{prodone}), we find
\be
\label{gvhet}
\sum_{r \in {\rm Pic (K3)}}\sum_{g=0}^{\infty}
n_g(r) z^{g} q^{r^2/2} =f(q) \xi^2(z),
\ee
where $z=4 \sin^2 (\lambda/2)$, and
$\xi(z)$ is the function that appears in helicity supertraces (see for example \cite{kiritsis,ddmp})
\be
\label{helicityxi}
\xi(z)=\prod_{n=1}^{\infty} {(1-q^n)^2 \over 1-2 q^n \cos \lambda  + q^{2n}}=\prod_{n=1}^{\infty} {(1-q^n)^2 \over (1-q^n)^2 + z q^n}.
\ee
A similar expression was written down in \cite{hosono} in a particular example. Equation (\ref{gvhet}) applies to many different
heterotic duals, like the ones studied in \cite{mm,kkrs}. Notice that, if the modular form $f(q)$ has an expansion in $q$ with integer coefficients, then integrality of $n_g(r)$ is manifest.
The final result for the generating functional of the $n_g(r)$ involves a model-dependent quantity (the modular form $f(q)$) as
well as the universal factor $\xi^2(z)$. Therefore, in these heterotic models, the enumerative information of the $F_g$s is
encoded in a single modular form $f(q)$, and this leads to a powerful principle which can be used to determine
these couplings in a
variety of models \cite{kkrs}.

\sectiono{The FHSV model}
\label{sec:heterotic}
In this section we will introduce and study the FHSV model of \cite{fhsv}. We will first discuss the heterotic side and
give some details about the one-loop partition function which will be needed in the computation of the
$F_g$ couplings. Then we discuss the type IIA side and the geometry of the Calabi-Yau, which will be
important to give an interpretation of the couplings and in the B-model analysis of section 6.

\subsection{The heterotic side of the FHSV model}
The FHSV model is defined, on the heterotic side, by an asymmetric orbifold \cite{fhsv}. One first considers
the splitting of the compactification lattice $\Gamma^{6,22}$ as
\be
\Gamma_u=\Gamma^{1,9}_1 \oplus \Gamma^{1,9}_2\oplus  \Gamma_s^{1,1} \oplus  \Gamma^{2,2}\oplus \Gamma_g^{1,1},
\label{eq:latticeheterotic}
\ee
where each of the $\Gamma^{1,9}$ can be further decomposed as
\be
\Gamma^{1,9}=\Gamma^{1,1}_d \oplus E_8(-1).
\ee
We now act with a $\ZZ_2$ symmetry as follows:
\be
|p_1, p_2, p_3, p_4, p_5\rangle \rightarrow e^{ \pi i \delta\cdot p_3}|p_2, p_1,p_3, -p_4, -p_5\rangle
\label{eq:orbifoldaction}
\ee
where $\delta =(1,-1) \in \Gamma^{1,1}_s$, and $\delta^2=-2$. Therefore $\ZZ_2$
acts as an exchange symmetry in the direct sum $\Gamma^{1,9}_1 \oplus \Gamma^{1,9}_2$, as a
shift in $\Gamma_s^{1,1}$, and as $-1$ in $\Gamma^{2,2}\oplus \Gamma_g^{1,1}$.  It is easy to see \cite{fhsv}
that this asymmetric orbifold leads to an heterotic string compactification with ${\cal N}=2$ supersymmetry in
four dimensions. The massless spectrum consists of 11 vector multiplets, 11 hypermultiplets, and the supergravity
multiplet.

The vector multiplet moduli space for this compactification is given by
\be
{\rm SL}(2, \ZZ) \backslash {\rm SL}(2, \IR)/SO(2) \times {\cal M},
\ee
where
\be
\label{vmms}
{\cal M}=O(\Gamma_1)\backslash O(2,10)/[O(2)\times O(10)],
\ee
and $O(\Gamma_1)$ is the group of automorphisms of the lattice
\be
\Gamma_1= \Gamma^{1,1}_s \oplus \Gamma_d^{1,1}(2) \oplus E_8(-2).
\ee
This is in fact the lattice associated to the untwisted, projected sector of the orbifold.

As a warm up exercise, we will now compute the one-loop partition function of the FHSV orbifold,
since the results will be useful for the computation of the $F_g$ amplitudes (the helicity supertrace generating
function of this model has been independently computed in the recent paper \cite{ddmp}).
We will denote by $Z[{h\atop g }]$ the partition
functions on the sector twisted by $h$ and with the $g$ element inserted. Here, $g,h=0,1$ in the
usual way.

Let us first consider the bosonic sector. In the untwisted, unprojected sector we simply have
\be
Z^b[_0^0]= {1\over 2 {\bar \eta}^{24}(\tau) \eta^8(\tau)}
{\overline \Theta}_{\Gamma^u}(\tau).
\ee
In order to consider the other sectors, we introduce the lattices $\Gamma_J$ with
$J=1,2,3$:
\be
\label{lattices}
\Gamma_J=\Gamma^{1,1}_s \oplus \Gamma^{1,1}_d (\zeta_J) \oplus E_8(-\zeta_J),
\ee
The values of $\zeta_J$, $\alpha_J$, $\beta_J$ are given in table 1. The three different cases
$J=1,2,3$ correspond respectively to the orbifold blocks $01$, $10$ and $11$.
\begin{table}
\begin{center}
\begin{tabular}{|| l | r| r|r ||}
\hline
$J$ & 1 &2 &3\\ \hline
$\zeta_J$ & $2$ & $1/2$ & $1/2$ \\ \hline
$\alpha_J$  & $\delta$ & 0 & $\delta$\\ \hline
$\beta_J$  & 0 &$\delta$ & $\delta$ \\ \hline
\end{tabular}
\caption{$\zeta_J$, $\alpha_J$ and $\beta_J$ for the different blocks}\end{center}
\label{tablat}
\end{table}
In the untwisted, projected sector we identify the two sets of bosonic
excitations associated to the two $\Gamma^{1,9}$ lattices. This amounts to a doubling of the $\tau $ parameter in the
nonzero modes \cite{ks}. We then find,
\be
Z^b[^0_1]=
{4 \over {\bar \eta}^{9}(2 \tau) \eta(2\tau) {\bar \eta}^3(\tau){\eta}^3(\tau) }
\biggl| {\eta(\tau) \over \vartheta[^1_0](\tau)}\biggr|^3
{\overline \Theta}_{\Gamma_1}(\tau,\delta,0).
\ee
For the 10 and 11 orbifold blocks we find
\be
\begin{aligned}
Z^b[^1_0]=&{4\over {\bar \eta}^{9}(\tau/2) \eta(\tau/2){\bar \eta}^3(\tau){\eta}^3(\tau)}
\biggl| {\eta(\tau) \over \vartheta[^0_1](\tau)}\biggr|^3
{\overline \Theta}_{\Gamma_2}(\tau, 0, \delta),\\
Z^b[^1_1]=&
{4\over {\bar \eta}^{9}({\tau +1\over 2}) \eta({\tau+1\over 2}){\bar \eta}^3(\tau){\eta}^3(\tau)}
\biggl| {\eta(\tau) \over \vartheta[^0_0](\tau)}\biggr|^3
{\overline \Theta}'_{\Gamma_3}(\tau, \delta, \delta).
\end{aligned}
\ee
In the 11 block, the $'$ in the theta function indicates that the sum over
lattice vectors includes an insertion of
\be
\label{signinsert}
(-1)^{v^2},
\ee
where $v$ is the projection of $p$ onto $\Gamma^{1,1}({1\over 2})\oplus E_8(-{1\over2})$.

Let us now consider the fermionic sector in detail. The fermions in the $\Gamma_s^{1,1}$ lattice do not change under the $\ZZ_2$
symmetry, so together with the fermions in the uncompactified
directions we have
\be
Z^f_{\Gamma_s^{1,1}}[^a_b]=\biggl( {\vartheta [ {a\atop b}](\tau) \over
\eta (\tau)}\biggr)^{3/2}.
\ee
The orbifold blocks for
two complex fermions with symmetry $\psi
\rightarrow -\psi$ are given by (see for example \cite{kiritsis}, eq. (12.4.15)):
\be
 {\vartheta[^{a+h}_{b+g}] (\tau)\vartheta[^{a-h}_{b-g}] (\tau)
\over \eta^2}
\ee
Therefore, for the fermions in $\Gamma^{2,2}\oplus \Gamma_g^{1,1}$ one finds
\be
Z^f_{\Gamma^{2,2}\oplus \Gamma_g^{1,1}}[^h_g][^a_b]=
\biggl( {\vartheta[^{a+h}_{b+g}] (\tau)\vartheta[^{a-h}_{b-g}] (\tau)
\over \eta^2} \biggr)^{3/4}.
\ee
The treatment of the two fermions coming from $\Gamma^{1,9} \oplus \Gamma^{1,9}$
is slightly more delicate. The $00$ block in the $a,b$ sector is simply
\be
Z^f_{\Gamma^{1,9} \oplus \Gamma^{1,9}}[_0^0][^a_b]={\vartheta [^a_b](\tau) \over
\eta (\tau)}.
\ee
Let us now analyze the invariant states in the NS sector. A convenient basis for the Hilbert space
${\cal H}_{\rm NS}^{(1)} \otimes {\cal H}_{\rm NS}^{(2)}$ is given by
\be
\begin{aligned}
& \biggl( \psi^{(1)}_{-n_1} \cdots \psi^{(1)}_{-n_{2k}}\psi^{(2)}_{-m_1} \cdots \psi^{(2)}_{-m_l} \pm (1\leftrightarrow 2)\biggr) |0\rangle, \\
& \biggl( \psi^{(1)}_{-n_1} \cdots \psi^{(1)}_{-n_{2k+1}}\psi^{(2)}_{-m_1} \cdots \psi^{(2)}_{-m_{2l+1}} \mp (1\leftrightarrow 2)\biggr) |0\rangle,
\end{aligned}
\ee
where $n_i, m_i >0$ are half-integers. The above states
have the sign $\pm 1$, respectively, under the $\ZZ_2$ symmetry generator $g$ which
exchanges the two lattices. It is easy to see that in computing the trace over the
Hilbert space with an insertion of $g$, the above states cancel except when the $(1)$ and the $(2)$ content
is the same. Therefore, only the states
\be
\begin{aligned}
&
\psi^{(1)}_{-n_1} \cdots \psi^{(1)}_{-n_{2k+1}}\psi^{(2)}_{-n_1} \cdots \psi^{(2)}_{-n_{2k+1}}|0\rangle,\\
& \psi^{(1)}_{-n_1} \cdots \psi^{(1)}_{-n_{2k}}\psi^{(2)}_{-n_1} \cdots \psi^{(2)}_{-n_{2k}}|0\rangle
\end{aligned}
\ee
contribute to the trace, with signs $-1$ and $+1$ under $g$, respectively. An odd number of fermion oscillators
leads to a $-1$ sign, but this is like having an insertion of $(-1)^F$. We then find
\be
{\rm Tr}_{{\cal H}_{\rm NS}^{(1)} \otimes {\cal H}_{\rm NS}^{(2)}} g \, q^{L_0 -c/24}={\rm Tr}_{{\cal H}_{NS}} (-1)^F \, q^{2L_0 -c/12}=
\biggl( {\vartheta [_1^0](2\tau) \over
\eta (2\tau)}\biggr)^{1\over 2},
\ee
where the doubling in $\tau$ is due to the doubling in the oscillator content. Notice that the insertion of $(-1)^F$ in the above trace does
not change anything, since $(-1)^{F_1}$ and
$(-1)^{F_2}$ cancel each other, therefore
\be
Z^f_{\Gamma^{1,9} \oplus \Gamma^{1,9}}[^0_1][^a_b] =
\biggl( {\vartheta [^a_1](2\tau) \over
\eta (2\tau)}\biggr)^{1\over 2},
\ee
and the expressions for the other blocks can be obtained by modular
transformations.
%
%

Putting all these results together, we can write up the one-loop partition functions for the different blocks.
One finds, for example:
\be
Z[_0^0]={1\over 2 {\bar \eta}^{24}(\tau) \eta^8 (\tau)}{\overline \Theta}_{\Gamma_u}(\tau)
 \sum_{a,b} (-1)^{a + b + ab}
\biggl( {\vartheta \bigl[^a_b\bigr] (\tau)\over \eta(\tau)}\biggr)^4.
\ee
for the $00$ block. For the $01$ block, one finds
\be
\label{invariant}
\begin{aligned}
Z[^0_1]=&{4 \over  {\bar \eta}^9 (2\tau) \eta (2\tau) |\eta (\tau)|^3} {1\over
|\vartheta \bigl[{1\atop 0}\bigr] (\tau) |^3 }{\overline \Theta}_{\Gamma_1}(\tau, \delta,0)\\
&\times { \bigl(\vartheta [{0\atop 0}] (\tau)\bigr)^{3/2}
 \bigl(\vartheta [{0\atop 1}] (\tau)\bigr)^{3/2} \over \eta^3(\tau) }{\bigl( \vartheta [^0_0](2 \tau) \bigr)^{1/2}
-\bigr( \vartheta [{0\atop 1}](2 \tau)\bigr)^{1/2} \over \bigl( \eta(2 \tau)\bigr)^{1\over2}}.
\end{aligned}
\ee
\subsection{The type II side of the FHSV model}
\label{sec:fhsvgeom}

The dual type II realization of the FSHV model is a compactification on the Enriques
CY $M$ with holonomy $SU(2)\times \ZZ_2$. The two covariant constant spinors of
opposite chirality on $M$ lead to ${\cal N}=2$ supersymmetry in four dimensions, but many features of
the model are between the $SU(2)$ holonomy case with ${\cal N}=4$ supersymmetry and
the generic situation with $SU(3)$ holonomy and ${\cal N}=2$ supersymmetry.

The compactification manifold $M$ of the type II string is constructed
as a free quotient of the manifold $Y={\rm K3}\times \bT^2$. The $\ZZ_2$ acts as the
free Enriques involution~\cite{bpv} on the K3 and as inversion
$\ZZ_2:z\mapsto -z$ on the coordinate $z$ of the $\bT^2$.
If $\bT^2=\IC/\ZZ^2$ is defined by $z\sim z+1\sim z+\tau$, we have four $\ZZ_2$ fixed points at
$\{p_1,p_2,p_3,p_4\}=\{0,{1\over 2},{\tau\over 2}, {1\over 2}+{\tau\over 2}\}$.
The geometry of the $\bT^2/\ZZ_2$ orbifold is that of a $\IP^1$ with area
${\rm Im}(\tau)/2$ and four conical curvature singularities at the $p_i$
each of which has deficit angle $\pi$. The total space $M$ is a K3 fibration
over the $\IP^1$, and by construction it has Enriques fibres $E$ of multiplicity two,
over the four $p_i$\footnote{It also exhibits an elliptic fibration over the Enriques
$E$ surface with four sections. An Enriques surface itself has two elliptic fibres with
multiplicity two and can be obtained  from $dP_9$ --a $\IP^2$ blown up in nine points-- by
a logarithmic transform on two fibres.}.

Every Enriques surface $E={\rm K3}/\ZZ_2$ is a free quotient of a K3 by the Enriques
involution $\rho:{\rm K3} \rightarrow {\rm K3}$. In order to construct a type II realization of the FHSV model, one first notices that the
two-cohomology lattice $H^2({\rm K3},\ZZ)$,
\begin{equation}
\Gamma_{\rm K3}=\Gamma^{3,19}=\Gamma^{1,9}_1\oplus \Gamma^{1,9}_2\oplus \Gamma^{1,1}_g
\label{g193}
\end{equation}
can be identified with the same
blocks that appear in (\ref{eq:latticeheterotic}). The Enriques involution
$\rho^*(p_1\oplus p_2\oplus p_5)=p_2\oplus p_1\oplus (-p_5)$ on
$\Gamma_{\rm K3}$ acts as in (\ref{eq:orbifoldaction}). The $\Gamma^{1,1}_s$
lattice is spanned by $H^0({\rm K3},\ZZ)$ and $H^4({\rm K3},\ZZ)$ in the type II
realization, and after quotienting by the involution $\rho$ it can be identified as
\be
\label{gammaones}
\Gamma^{1,1}_s =H^0(E,\ZZ) \oplus H^4(E,\ZZ).
\ee
The shift on this lattice in the orbifold (\ref{eq:orbifoldaction})
corresponds to turning on a Wilson line expectation value for the RR $U(1)$ fields \cite{aspinwall}.


Some properties of the model are most clearly seen in an algebraic realization.
We realize the two-torus as a hyperelliptic branched twofold
covering of $\IP^1$, with homogeneous coordinates denoted by $w:x$,
and described by the equation
\begin{equation}
y^2=f_4(w:x).
\label{elliptic}
\end{equation}
The $\ZZ_2$ acts as $\kappa:y\mapsto -y$, and the fixed points are the four branch
points $p_i$ of the degree four polynomial $f_4(w:x)=0$. The holomorphic
$(1,0)$ form
\be
\omega_{1,0}={{\rm d} x \over y}
\ee
is anti-invariant.

A similar
realization of a K3 admitting the Enriques involution $\rho$ is as a
double covering of $\IP^1\times \IP^1$ branched at the vanishing locus of a bidegree
$(4,4)$ hypersurface in $\IP^1\times \IP^1$ \cite{bpv,horikawa}. The total
space is a eighteen-parameter family of K3 surfaces
\begin{equation}
{\cal Y}^2=f_{4,4}(s:t,u:v)\ .
\end{equation}
The Enriques involution acts freely as
\be
\rho:({\cal Y},s:t,u:v)
\mapsto (-{\cal Y},s:(-t),u:(-v))
\ee
on a symmetric but otherwise generic  slice of the
family. The holomorphic $(2,0)$ form is given by
\be
\omega_{2,0}={s u {\rm d} t
\wedge {\rm d} v\over {\cal Y}}.
\ee
Since $\rho$ acts freely, the
fundamental group of the Enriques surface $E$ is $\ZZ_2$ and the Euler number
is $\chi(E)=\chi({\rm K3})/2=12$.  As $\omega_{2,0}$ (and $\bar \omega_{2,0}=\omega_{0,2}$)
is anti-invariant, the cohomology groups have dimensions $h^{00}=h^{22}=1$,
$h^{10}=h^{01}=h^{20}=h^{02}=0$ and $h^{11}=\chi(E)-2=10$.
The canonical bundle is a two
torsion class, i.e. $K^{\otimes 2}_E ={\cal O}_E$, hence non trivial: $K_E\neq {\cal O}_E$.
On the blow up of the special configuration with
\be
f_{4,4}=(u-v)(u+v)(a s^4(u^2-v^2) +
b s^2 t^2 ( u^2-v^2)+ t^4 ( c u^2 + d v^2))
\ee
and with Picard number 18, one can explicitly check~\cite{bpv}
that the invariant part $\Gamma_{\rm K3}^+$ and anti-invariant part $\Gamma_{\rm K3}^-$ of $\Gamma^{3,19}$ under $\rho^*$ are
\be
\Gamma_{\rm K3}^+=\Gamma^{1,1}(2)\oplus E_8(-2), \qquad \Gamma_{\rm K3}^-=[\Gamma^{1,1}(2) \oplus  E_8(-2) ]\oplus
\Gamma_g^{1,1}.
\ee
The middle cohomology $H^2(E,\ZZ)$ is isometric to the lattice
\be
\label{twoelattice}
\Gamma_E={1\over 2}\Gamma_{\rm K3}^+=\Gamma^{1,1}(1)\oplus E_8(-1).
\ee

The Calabi-Yau manifold $M$ is constructed as $M=({\rm K3}\times \bT^2)/\ZZ_2$, where the $\ZZ_2$ acts as
\be
(\rho,\kappa):({\cal Y},s:t,u:v,y,w:x) \mapsto (-{\cal Y},s:
-t,u:-v,-y:w:x).
\ee
On the generic K3 fiber the $\ZZ_2$ acts as the monodromy
$\rho$, when the corresponding base point is transported in a loop around
the special points $p_i$. The $\ZZ_2$ part of the holonomy is generated as follows. A tangent vector
$v\in T_M$ transported over a nontrivial loop in the base (\ref{elliptic})
around two points of total deficit angle of $\pi$ is inverted: $v\mapsto -v $.
In the base direction the inversion occurs because of the deficit angles, and
in the fiber direction due to the monodromy $\rho$.

The cohomology of $M$ is easy to find. $\Omega=\omega_{2,0}\wedge \omega_{1,0}$ is
invariant and becomes the unique, nowhere vanishing $(3,0)$-form on $M$.
The $10$ invariant $(1,1)$ forms $\omega_{1,1}^{(i)}$, $i=1,\ldots, 10$
in $\Gamma_{\rm K3}^+$, together with the volume form on $\IP^1$, $\omega_{1,1}$,
give $11$ harmonic forms in $H^{1,1}(M,\ZZ)$. We will
adopt the type IIA interpretation in which the vector multiplets are
mapped to the complexified K\"ahler moduli.  Notice that the heterotic moduli of the
Narain compactification are mapped to the K\"ahler moduli of the fiber (as we will make explicit in the next section),
while the heterotic dilaton $S$ is mapped to the
complexified K\"ahler modulus of the $\IP^1$ base. Since $\chi(M)=0$,
one has $h^{2,1}=11$. Ten of these forms can be explicitly constructed by taking the ten forms in $\Gamma_{\rm K3}^-$ of type $(1,1)$ and
forming their wedge product with $\omega_{1,0}$. The remaining $(2,1)$ form is $\omega_{2,0}\wedge \omega_{0,1}$.

The moduli space of $M$ has two different types of singular loci \cite{fhsv,aspinwall}
which lead to conformal field theories in four dimensions.
The first degeneration comes from the shrinking of a smooth rational
curve $e\in \Gamma_E$ with $e^2=-2$. Since $\IP^1$ has no unramified
cover, the preimage of $e$ in $\Gamma_{\rm K3}$ must be the sum $e_1+e_2$ of two spheres $e_1$, $e_2$ in $\Gamma_{\rm K3}$
with $e_i \in \Gamma^{1,9}_i$ and $\rho^*(e_1)=e_2$. If $e$ goes to zero size so do $e_1$ and $e_2$ in $\Gamma_{\rm K3}$.
The shrinking $\IP^1$ leads to an $SU(2)$ gauge symmetry enhancement:
in type IIA  theory, a D2-brane wrapping the $\IP^1$ with two possible orientations
leads to massless $W^\pm$ bosons, which complete the corresponding
$U(1)$ vector multiplet to a vector multiplet in the adjoint representation.
This is plainly visible in the spectrum of the perturbative heterotic string,
where the gauge group is realized by a level
$2$ WZW current. For each vanishing $e_1+e_2$ in $\Gamma_{\rm K3}^+$ there is a vanishing $e_1-e_2$
in the first summand of $\Gamma_{\rm K3}^-$ which leads to a hypermultiplet, also in
the adjoint representation of the gauge group. We then obtain for
this point the massless spectrum of ${\cal N}=4$ supersymmetric gauge theory,
which has a vanishing beta function and no Higgs branch.

The second degeneration is again plainly visible in the
perturbative heterotic string and arises if one goes to
the selfdual point in the lattice $\Gamma_s^{1,1}$
factor in (\ref{eq:latticeheterotic}). As usual one gets
a $SU(2)$ gauge symmetry enhancement at level $1$. In addition
one gets four hypermultiplets in the fundamental representation
of $SU(2)$, one from each fixed point of the $\bT^2$. The resulting gauge theory
is ${\cal N}=2$, $SU(2)$ Yang-Mills theory with four massless hypermultiplets. This
theory has a vanishing beta function and it is believed to be conformal \cite{swtwo}. It
also has a Higgs branch which leads to a transition to a generic simply connected CY
with $SU(3)$ holonomy and Hodge numbers $h_{21}=10$
and $h_{11}=16$ \cite{fhsv,aspinwall}.

An interesting difference between the two degenerations is that the first
one occurs when a two-cycle of the covering K3 becomes small,
while in the second one the full K3 surface has a volume of order the Planck
scale \cite{aspinwall}. 

The fact that these degenerations are associated to
conformal theories indicates that there are no genus zero
contributions to the Gromow-Witten invariants in type IIA theory on $M$ \cite{fhsv}.
For these degenerations, the K\"ahler class of the base is identified
with the scale of the gauge coupling constant. An eventual scale
dependence in ${\cal N}=2$ supersymmetric theories comes from a one-loop
correction to the beta function, which corresponds to worldsheet instantons
with degree zero in the base, and from space time instantons, which are
in turn related to the growth of worldsheet instantons with non-vanishing degree in the
base. Both contributions are expected to vanish for the conformal theories. Later we
will check with explicit computations that indeed there are no
worldsheet instanton corrections to the type IIA prepotential\footnote{A direct A-model
argument can be given using relative Gromov-Witten invariants \cite{mp}.}.

\sectiono{Heterotic computation of the $F_g$ couplings}
\label{sec:fg}

In this section we compute the couplings $F_g$ in the heterotic side. It turns out that there
are two natural lattice reductions to perform the computation: the geometric reduction, and
the Borcherds-Harvey-Moore (BHM) reduction. We will present the results for the couplings in both
reductions and we will also propose a type IIA interpretation of
these results.

Before doing the lattice reduction, we have to evaluate the integrand (\ref{Fgint}) for the heterotic
FHSV model. This is rather straightforward by using the
results of the previous section. We have four orbifold blocks, but the
first block (corresponding to $h=g=0$) vanishes. The blocks $(h,g)=(0,1), (1,0), (1,1)$ will be labelled by
$J=1,2,3$, and an easy computation shows that the modular forms $f_J(q)$ in (\ref{gintegrand}) are given by
\be
\label{fhsvforms}
\begin{aligned}
f_1(q)&=-{128\over \eta^6(\tau) \vartheta_2^6 (\tau)}=-{2 \over q} \prod_{n=1}^{\infty} (1-q^{2n})^{-12},\\
f_2(q)&={4\over \eta^6(\tau) \vartheta_4^6 (\tau)}=4q^{-{1\over 4}}\prod_{n=1}^{\infty}(1-q^{n})^{-12}
(1-q^{n-1/2})^{-12},\\
f_3(q)&={4\over \eta^6(\tau) \vartheta_3^6 (\tau)}=4 q^{-{1\over 4}}\prod_{n=1}^{\infty}(1-q^{n})^{-12}
(1+q^{n-1/2})^{-12}.
\end{aligned}
\ee
The Narain lattices for $J=1,2,3$ are given in (\ref{lattices}), and the corresponding
theta functions in (\ref{gintegrand}) are the same ones that appear in the computation
of the one-loop partition function in the previous section. The modular forms in (\ref{fhsvforms}) have the right modular
weight: the conjugate Narain-Siegel theta function for a lattice of signature $(2,10)$ with $2g-2$ insertions
has modular weight $(5,2g-1)$, the modular form ${\widehat \CP}_g$ has modular weight $(2g,0)$, and
the insertion $\tau_2^{2g-1}$ has modular
weights $(-2g+1,-2g+1)$. Taking into account that the weight of the forms $f_J(q)$ is $(-6,0)$, we see
that the integrand in (\ref{Fgint}) has zero modular weight, as it should. Notice that the Narain-Siegel theta functions
involve nonzero $\alpha, \beta$ and lattices which are not
self-dual. This means that we will have to modify in an appropriate way the computation in \cite{mm}.
We will now present the computation
of the couplings in both reductions.

\subsection{The geometric reduction}
\label{sec:geometricreduction}

In order to apply the reduction technique we need explicit formulae for the
projections of the lattice, that in turn depend on the moduli. Let us write a vector
$p \in \Gamma_J$ as
\be
\label{pvector}
p=(n, m, n_2, m_2, \vec q),
\ee
where $(n, m)$, $(n_2, m_2)$ are integer coordinates on $\Gamma^{1,1}_s$ and $\Gamma^{1,1}_d (\zeta_J)$, and
$\vec q$ is a vector of integer coordinates in the $E_8$ lattice.
The norm of $p$ is given by
\be
\label{pnorm}
p^2=p_R^2 - p_L^2= 2nm + \zeta_J (2 n_2 m_2 - \vec q^{~2}).
\ee
Since the lattices $\Gamma_J$ have two $\Gamma^{1,1}$ factors, there are two
natural reductions that one can use.  The first one will be referred to as the geometric reduction. The reason for this name is that, as we will see,
this reduction leads to an expression for $F_g$ which is valid in the
large volume limit of the K\"ahler moduli space and gives a generating functional of Gromov-Witten invariants, or
equivalently, of BPS invariants that count D2-D0 bound states. In the geometric reduction, one chooses the reduction vector
\be
z=(1,0) \in \Gamma_s^{1,1}.
\label{geomz}
\ee
We then have $z'=(0,1) \in \Gamma_s^{1,1}$. The reduced lattice is
\be
K_J=E_8(-\zeta_J) \oplus \Gamma_d^{1,1}(\zeta_J), \qquad J=1,2,3.
\ee
Different choices of reduction vectors correspond to different
choices of cusps in the moduli space, and in particular lead to different parameterizations of this space. To make this
explicit, we remind that the exact moduli space of vector multiplets for the K\"ahler parameters of the
fiber is the coset $O(2,10)/[O(2)\times O(10)]$.
This coset is given by the following algebraic equations
satisfied by the complex variables $(w_1, \cdots, w_{12})$
\cite{fklz,cardoso}
\be
\begin{aligned}
\sum_{i=1}^{10} |w_i|^2 - |w_{11}|^2 - |w_{12}|^2=&-2 Y,  \\
\sum_{i=1}^{10} w_i^2 - w_{11}^2 - w_{12}^2=&0.
\end{aligned}
\ee
The quantity $Y$ above is the same one that appears in (\ref{gintegrand})
and the mass formula gives
\be
|p_R|^2={|v\cdot w|^2 \over Y},
\ee
where the vector $v$ is defined by
\be
v=\biggl( \zeta_J^{1\over 2} \vec q, m -{1\over 2} n, \zeta_J^{1\over 2} (m_2 -{1\over 2} n_2),
m +{1\over 2} n, \zeta_J^{1\over 2} (m_2 +{1\over 2} n_2)\biggr).
\ee
For the reduction vector (\ref{geomz}) it is convenient to parameterize the coset by ten independent complex coordinates
\be
\label{parameters}
y=(y^+, y^-, \vec y),
\ee
which are defined as follows
\be
\label{geompar}
\begin{aligned}
w_j=&y_j,  \quad j=1, \cdots, 8,\qquad
w_9=1+{1\over 4} y^2, \\
w_{10}=&{1\over 2} (y^+-2 y^-),\qquad
w_{11}=-1+{1\over 4} y^2 ,\\
w_{12}=&{1 \over 2} (y^++ 2y^-),
\end{aligned}
\ee
where the (complex) norm of the vector (\ref{parameters}), $y^2$, is defined by
\be
y^2= 2 y^+ y^- - \vec y^{~2}.
\ee
We will denote by $y^{\pm}_2$, $\vec y_2$ the imaginary parts of these moduli. From the above
parameterization one finds,
\be
Y=\bigl( {\rm Im}\, y\bigr)^2= 2 y_2^+y_2^--\vec y_2^{~2},
\ee
and $p_R =v\cdot w/Y^{1\over 2} \in {\bC}$
is given by
\be
\label{prsqgeom}
p_R={1\over Y^{1/2}}\biggl(-n +{1\over 2} m y^2 +\zeta_J^{1\over 2}  m_2  y^+ +\zeta_J^{1\over 2}  n_2y^- +\zeta_J^{1\over 2}\vec q \cdot \vec y \biggr).
\ee
With this parameterization, the resulting topological couplings will have good convergence properties in the region ${\rm Im}\, y \rightarrow \infty$.
Notice that
\be
|z_+|={1\over Y^{1\over 2}},
\ee
therefore $\nu=1$. It is easy to evaluate the exponent of (\ref{oneargument}), which in this case it is equal to
\be
\label{cousa}
2\pi \ri r\cdot y=2 \pi \ri \zeta_J^{1\over 2} \Bigl( m_2  y^+ +n_2y^- +\vec q \cdot \vec y \Bigr),
\ee

We can now proceed to evaluate the integral (\ref{Fg}). The first thing to observe is that, with the choice
of reduction vector (\ref{geomz}), only the untwisted sector $J=1$ contributes. The reason for that is that in the twisted
sectors $J=2,3$, the lattice $\Gamma_s^{1,1}$ where the reduction is performed has the shift $\beta_J=\delta=z- z'$.
As shown in section 5 of \cite{mw},
in those cases the integral over the fundamental domain is zero. This is easy to understand by looking at the expression (\ref{thetared})
in Appendix B. The effect of this nonzero $\beta$ is to shift $c\rightarrow c-1/2$. As this integral is effectively evaluated
when $|z_+| \rightarrow 0$, the integrand vanishes. On the other hand, the theta function associated to
$\Gamma_1$ in the untwisted, projected sector $J=1$ includes a phase $e^{\pi i \delta\cdot p}$. It was shown in
\cite{lm} that the effect of this
phase is to shift the integer $\ell$ in (\ref{thetared}) as $\ell \rightarrow
\ell-{1\over 2}$. This means that the polylogarithm in (\ref{fgex}) becomes
\be
{\rm Li}_{m}(x) =\sum_{\ell=1}^{\infty} {x^{\ell}\over \ell^m} \rightarrow
\sum_{k=0}^{\infty} {x^{k +{1\over2}}\over (k +{1\over 2})^m} =
2^m {\rm Li}_m (x^{1\over 2}) - {\rm Li}_{\rm m}(x),
\label{lishift}
\ee
with $m=3-2g$. In order to write the argument of the polylogarithm, we will relabel $m_2, n_2 \rightarrow m, n$, and introduce the moduli parameters $t=(t^+, t^-, \vec t)$ as
\begin{equation}
-2\pi \ri y ={\sqrt 2}t.
\end{equation}
The integer vector $r$ in (\ref{cousa}) reads
\be
\label{rlattice}
r=(n,m,\vec q) \in \Gamma_E=\Gamma^{1,1} \oplus E_8(-1),
\ee
and it follows from (\ref{twoelattice}) that it labels two-cohomology classes in the Enriques fiber. It has norm $r^2=2mn -\vec q^{~2}$. 
The argument of the polylogarithm is
$$
x^{1\over 2}=\exp(-r\cdot t ), \qquad r\cdot t= mt^+ + nt^- + \vec q \cdot \vec t.
$$
The final formula for the $F_g$ is then
\be
F_g(t)=\sum_{r>0} c_g(r^2) \bigg\{ 2^{3-2g} {\rm Li}_{3-2g}(\re^{-r\cdot t}) - {\rm Li}_{3-2g}(\re^{-2 r\cdot t})\biggr\},
\label{heteroticprediction}
\ee
where
\be
\label{geomrmod}
\sum_n c_g(n) q^n =f_1(q){\cal P}_{g}(q),
\ee
$f_1(q)$ is given in (\ref{fhsvforms}). The restriction $r>0$ means that \cite{hm} $n>0$, or $n=0, m>0$, or $n=m=0$, $\vec q>0$.
The reason that the above formula involves $c_g(r^2)$ instead of $c_g(r^2/2)$ as in (\ref{fgex}) is simply that
the norm of the reduced lattice is twice the norm of the lattice in (\ref{rlattice}). Notice that due to the shift
in $\ell$ there is no contribution from the ``zero orbit'' (\ref{degbor}). The above expression is only valid in principle for
$g>0$, and the computation of the prepotential involves a somewhat different procedure explained in \cite{hm}. It is easy to check
however that the worldsheet instanton corrections to the prepotential
are given by (\ref{heteroticprediction}) specialized to $g=0$ (the same thing happens in
the STU model analyzed in \cite{mm}). Since $r^2$ is always even and $f_1(q)$ has no even powers of $q$ in its expansion, we conclude that
the instanton corrections to $F_0$ vanish along the fiber directions. This is in agreement with the analysis of \cite{fhsv}.

The genus one amplitude can be written as follows:
\be
\label{anotherborcherds}
F_1(t) = -{1\over 2} \log \, \prod_{r>0} \biggl( { 1-\re^{-r\cdot t} \over 1+\re^{-r\cdot t} }\biggr)^{2 c_1(r^2)}.
\ee
The infinite product appearing in this equation was previously found by Borcherds in a related context 
\cite{borcherdstwo}. In the Example 13.7 of that paper, Borcherds considers 
two different expressions for the same automorphic form, obtained by expanding it around different cusps. 
Both expressions are denominator formulae for two different superalgebras. The second denominator formula in this Example 
is precisely the infinite product of (\ref{anotherborcherds}) (to see this one notices that the part of 
$2 f_1(q) {\cal P}_1(q)$ involving even powers of $q$ equals the modular form $f_{00}(2\tau)$ introduced by Borcherds). 

We now propose the following type IIA interpretation of this computation.
As shown in (\ref{gammaones}), the reduction lattice $\Gamma^{1,1}_s$ we are choosing here
corresponds to the $H^0(E, \ZZ) \oplus H^4(E, \ZZ)$ cohomology of
the Enriques surface, and $z$ and $z'$ are integer generators of $H^0(E, \ZZ)$ and $H^4(E, \ZZ)$, respectively. The
remaining lattice can be identified to $H^2(E, \ZZ)$, and indeed $r$ in (\ref{rlattice})
is a set of integer coordinates for two-homology
classes on the Enriques fiber. For this reduction, the
region of moduli space where ${\rm Im}\, y\rightarrow \infty$ is the region where the D2s and the D0 are light (as
shown in (\ref{prsqgeom}), their mass goes like $1/Y^{1\over2}$ and $y/Y^{1\over2}$, respectively) while the D4s are heavy
(their masses go like $y^2/Y^{1\over2}$). Therefore, the region ${\rm Im}\, y$ where
this reduction is appropriate is the region of moduli space where the D2 and D0 are the lighter states,
and the D4s wrapping the Enriques fiber are heavy. This is the
large volume limit, and we expect the answer for $F_g$ to encode information about Gromov-Witten
invariants of the Enriques fiber. The sum over $\ell$ in (\ref{fgex}) can
be interpreted as the Poisson resummation of a sum over D0 brane charges
(notice that $\ell$ appears after Poisson resummation of the integer $n$ in (\ref{prsqgeom})),
and the shift in (\ref{lishift}) corresponds to
the RR Wilson line background along the $H^0(E, \ZZ)$ direction \cite{fhsv,aspinwall}. To substantiate this
interpretation, we will see in the next section that (\ref{heteroticprediction}) matches
with the geometric computation of BPS invariants proposed in \cite{kkv}. Moreover, in section 6
we will find perfect agreement of the heterotic predictions with a B-model
computation of $F_g$ for $g\le 4$.

Although (\ref{heteroticprediction}) is similar to other results obtained for
heterotic models, it has some additional properties that make it particularly simple.
For example, one can show that
\be
\label{twoderfone}
C_{ij} {\partial^2 F_1 \over \partial t_i \partial t_j} = -16 F_2,
\ee
where $C_{ij}$ is the intersection matrix of $\Gamma^{1,1} \oplus E_8(-1)$.
This is a consequence of the following identity among the coefficients of the
modular forms (\ref{geomrmod}):
\be
n \, c_1(n)=-4 \, c_2(n), \quad n \, {\rm even},
\ee
which can be proved by comparing the even part of the
$\tau$ derivative of $f_1(q) {\cal P}_1(q)$ with the
even part of $f_1(q) {\cal P}_2(q)$.

We now discuss the other possible lattice reduction
available on the heterotic side.

\subsection{The BHM reduction}
\label{sec:borcherds}
Since the lattices (\ref{lattices}) include the sublattice $\Gamma^{1,1}_d(\zeta_J)$, it is natural
to compute the topological string amplitudes by choosing the reduction vector
\be
\label{borred}
z=(1,0) \in \Gamma^{1,1}_d (\zeta_J).
\ee
We call this the Borcherds-Harvey-Moore (BHM) reduction, since as we will see it is the choice of reduction made by Harvey and Moore
in \cite{hmborcherds}, and leads to the
infinite product introduced by Borcherds in \cite{borcherdsone}. The reduced lattice is then
$$
K_J=\Gamma^{1,1}_s \oplus E_8(-\zeta_J).
$$
Since we have chosen a different reduction vector, the associated parameterization of the moduli space will be
different from the one made in (\ref{geompar}). We introduce, as before, ten independent complex coordinates $y=(y^+,y^-, \vec y)$ defined
through:
\be
\label{bhmpar}
\begin{aligned}
w_j=&y_j,  \quad i=1, \cdots, 8,\qquad
w_9={1\over 2} (y^+ - 2 y^-), \\
w_{10}=&1+{1\over 4} y^2,\qquad
w_{11}={1\over 2} (y^++ 2y^-),\\
w_{12}=&-1+{1\over 4}y^2.
\end{aligned}
\ee
Although we have used the same notation for the $y$ coordinates, they are related to the $w$ coordinates in a different way than in the
geometric reduction. With this parameterization, we find
\be
\label{prsq}
p_R={1 \over Y^{1/2}}\biggl(-\zeta_J^{1\over 2}  n_2 +{1\over 2} \zeta_J^{1\over 2} m_2 y^2 +n y^- + my^+ +\zeta_J^{1\over 2}\vec q \cdot \vec y \biggr).
\ee
Notice that the reduction vector has the norm
\be
\label{rednorm}
|z_+|=\biggl( {\zeta_J \over Y}\biggr)^{1\over 2},
\ee
hence the quantity $\nu$ introduced in (\ref{zquot}) has the value $\nu=\zeta_J$ for each block.
The exponent of (\ref{oneargument}) is now
\be
\nu^{-{1\over 2}} 2\pi \ri r\cdot y=2 \pi \ri \zeta_J^{-{1\over 2}}\bigl( ny^-+ my^+ + \zeta_J^{1\over 2}  \vec q\cdot \vec y\bigr).
\ee
The computation of the integral (\ref{Fgint}) is very similar to the one performed in \cite{mm}, which we summarized in
section 2 and Appendix B. The answer in this case is a sum over the three orbifold blocks $J=1,2,3$, involving different lattices.
One has now to be careful with the effects of the shifts $\alpha$, $\beta$. Since with this choice of reduction vector the shifts are
orthogonal to $z,z'$, they only lead to insertions of phases in the sum over the reduced lattice,
as well as to shifts in their vectors, and their effect is easy to track. To write down the final answer, we first define the coefficients of modular forms:
\be
{\cal P}_{g}(q)f_J(q)=\sum_n c_g^J(n) q^n.
\ee
Then, the couplings $F_g$ are given by
\be
\label{fganswer}
\begin{aligned}
&F_g= 2^{1-g} \sum_{r>0} (-1)^{m+n} c_g^1 (mn-\vec q^{~2} ) {\rm Li}_{3-2g}(e^{-r\cdot t}) \\
&+2^{g-1}\sum_{r > 0} \biggl\{ c_g^2 \biggl({1 - \vec q^{~2} \over 4} + mn+ {m+n\over 2}\biggr) -(-1)^{\vec q^{~2}/2 + m +n}
c_g^3 \biggl({1-\vec q^{~2} \over 4} + mn+ {m+n\over 2}\biggr)\biggr\}\\
& \qquad \cdot {\rm Li}_{3-2g}(e^{-r\star t}).
\end{aligned}
\ee
In this equation, $r=(m, n,\vec q)$, the coordinate $t=(t^{\pm}, \vec t)$ is defined in terms of $y$ by
\be
-2\pi \ri y =({\sqrt 2} t^{\pm}, \vec t),
\ee
and the inner products in (\ref{fganswer}) are given by
\be
\label{inns}
\begin{aligned}
r\cdot t =&mt^+ + nt^- + \vec q \cdot \vec t, \\
r\star t =&(2m+1)t^+ + (2n +1)t^- + \vec q \cdot \vec t.
\end{aligned}
\ee
The insertions $(-1)^{m+n}$ in the first and last block are due to the nonzero $\alpha=\delta$, while the shift in the second inner product in (\ref{inns}) is
due to the shift by $\beta=\delta$, and we have relabelled $n \rightarrow n+1$.
Finally, the insertion of $(-1)^{\vec q^{~2}/2}$ in the $J=3$ orbifold block comes from the insertion (\ref{signinsert}).

The expression (\ref{fganswer}) can be simplified as follows. First, one notices that $c_g^2 (1/4 +p/2)$ equals $(-1)^{p+1}c_g^3(1/4+p/2)$. This is easy to see
by noting that they are the coefficients of modular forms related by $q^{1\over 2}\rightarrow -q^{1\over 2}$.
Therefore, the $J=2$ and $J=3$ contributions are equal and add up. We will call their contribution the contribution of the twisted sector, while the
contribution from $J=1$ will be called the contribution of the untwisted sector.
It is easy to see that the polylogarithms whose argument involves a K\"ahler class of
the form $m t^+ + nt^- + \vec q \cdot \vec t$ with $n$ and $m$ {\it both odd} receive contributions
from both the untwisted and twisted sector, while if $m$ or $n$ is even only the untwisted sector contributes. In the first case, the
contributions come from the coefficients of {\it odd} powers in the modular form
\be
\label{modgroup}
2^{1-g} {\cal P}_g (q) f_1(q) + 2^{g} {\cal P}_g(q^4) f_2(q^4),
\ee
while in the second case they come from the contributions of {\it even} powers in the first term in (\ref{modgroup}). However, since
the second term in (\ref{modgroup}) has only odd powers of $q$, we can use the modular form (\ref{modgroup}) for both cases.
As a last step, one notices by using doubling formulae (see Appendix A) that
\be
\label{modid}
{64 \over \eta^6(\tau) \vartheta_2^6 (\tau)} ={1\over  \eta^6(4 \tau) \vartheta_4^6 (4\tau)},
\ee
therefore $f_2(q^4)=-2 f_1(q)$. We can finally write down a compact expression for $F_g$ as follows:
\be
\label{finalfgbor}
F_g(t)=\sum_{r>0} c_g(r^2/2)(-1)^{n+m} {\rm Li}_{3-2g}({\rm e}^{-r \cdot t })
\ee
where the coefficients $c_g(n)$ are defined by
\be
\sum_n c_g (n) q^n =f_1(q)  \biggl\{ 2^{1-g} {\cal P}_{g}(q) -
2^{1+g} {\cal P}_{g}(q^4)\biggr\},
\label{finalmodcoef}
\ee
and in (\ref{finalfgbor}) we regard $r$ as a vector in $\Gamma^{1,1}\oplus E_8(-2)$, i.e. $r^2=2nm-2 \vec q^{~2} $.

We now consider some particular cases of (\ref{finalfgbor}) in more detail. Although the above expression is in principle valid for $g\ge 1$, one can see again that
the instanton corrections to the prepotential are given by its specialization to $g=0$, and
one finds $F_0=0$ due to a cancellation between the
untwisted and twisted sectors. Let us now look at $g=1$. This involves the modular form
${\cal P}_1(q)$ given in (\ref{casesps}). The doubling formulae (\ref{deis}) gives
\be
E_2(\tau)-4 E_2(4\tau)=-3 \vartheta_3^4(2\tau),
\ee
and one finds
\be
\label{sameborcherds}
F_1=-{1\over 2} \log \, \prod_{r>0} \Bigl(1- \re^{-r \cdot t}\Bigr)^{(-1)^{n+m} c_{\rm B}(r^2/2)}
\ee
where
\be
\sum_n c_{\rm B} (n) q^n ={\eta(2 \tau)^8 \over \eta(\tau)^8 \eta(4 \tau)^8}.
\ee
This is the modular form introduced by Borcherds in \cite{borcherdsone}, and the above
expression for $F_1$ agrees with that found by Harvey and Moore in \cite{hmborcherds} (up to a factor of $1/2$ due to
different choice of normalizations). The infinite product appearing in (\ref{sameborcherds}) is the denominator formula of 
a superalgebra, and it was pointed out in \cite{borcherdstwo} that it is actually identical to the infinite product in (\ref{anotherborcherds}), 
but expanded around a different cusp. This is of course expected, since in both cases we are evaluating the same integral, but 
with different choices of reduction vector. 

What is the interpretation of the $F_g$ amplitudes in the BHM reduction? In the remaining of this subsection,
we will make a proposal
for what is the enumerative content of the topological string amplitudes in this reduction. The first thing to notice
is that in the BHM reduction the reduced lattice
is $H^0(E, \ZZ)\oplus H^4(E, \ZZ)$ together with the $E_8(-2)$ sublattice of $H^2(E, \ZZ)$, and the integers $n$ and $m$ in
(\ref{finalfgbor}) label
zero and four-cohomology classes. The computation in this reduction is appropriate for the region ${\rm Im}\, y \rightarrow \infty$ in moduli space. However,
one can see from (\ref{prsq}) that this is the region where the light states are the D0, the D4 wrapping the
Enriques surface, the D2s in $E_8(-2)$, and one of the D2s in $\Gamma_d^{1,1}$, while the other D2 in this
sublattice (labelled by
$m_2$ in (\ref{prsq})) is heavy. Therefore,
we are {\it not} in the large radius regime, and the $F_g$ amplitudes
computed with this reduction do not have
an enumerative interpretation in terms of Gromov-Witten theory. It is easy to see that
they do not lead to integer GV invariants, and indeed
we will see in the next sections that the usual Gromov-Witten/D2-D0 counting interpretation has to be
reserved for the geometric reduction considered in the previous subsection. The $F_g$ couplings in the BHM reduction must
be counting bound states of the
light states associated to this cusp. One hint about their
BPS content comes from writing the generating functional as
\be
\label{bgen}
\sum_{g=0}^{\infty}\sum_{r>0} c_g(r^2/2)q^{r^2/2} \lambda^{2g-2}=f_1(q)\biggl\{{\xi^2(\lambda_+,q) \over 4 \sin^2( \lambda_+/2) } - 4 {\xi^2 (\lambda_-,q^4)\over
 4 \sin^2( \lambda_-/2) } \biggr\},
 \ee
 where $\xi(\lambda, q)$ is given in (\ref{helicityxi}), and
 \be
 \label{rescaled}
 \lambda_{\pm}= {\lambda\over 2^{\pm {1\over 2}}}.
 \ee
 The derivation of (\ref{bgen}) is very similar to the derivation of (\ref{gvhet}). This expression suggests that there should be a formula similar to (\ref{gvloop}),
 in which one does a Schwinger computation including the light states appropriate for this region of moduli space. The role of the D0s in the computation of
 \cite{gv} is now played by the light D2s in the reduction lattice. Since the BPS masses of these states have an extra factor $\nu^{1\over 2}$, the
 Schwinger integral in (\ref{gvloop}) becomes
 \be
 \label{gvmod}
 \begin{aligned}
& \int_0^{\infty} {ds \over s} \biggl( 2 \sin {s\over 2}\biggr)^{2g-2} \sum_{n_2=-\infty}^{\infty}
 \exp\Bigl[ -{s \over \lambda}(r\cdot y + 2\pi i\nu^{1\over2} n_2)\Bigr] =\\
 & \quad \sum_{\ell=1}^{\infty} {1\over \ell} \biggl( 2 \sin {\ell \lambda \over 2 \nu^{1\over 2} } \biggr)^{2g-2} \exp\Bigl[ -{\ell \over \nu^{1\over 2} }r\cdot y \Bigr],
 \end{aligned}
 \ee
where $n_2$ is the number of light D2 states. This fits with the heterotic expression given in (\ref{fgex}), and has the effect of rescaling the string coupling constant by
\be
\label{rescaling}
\lambda \rightarrow {\lambda\over \nu^{1\over 2}},
\ee
which explains the overall factor $\nu^{1-g}$ in (\ref{fgex}) as an effect of this rescaling, as well as the appearance of (\ref{rescaled}) in (\ref{bgen}).
In order for the logic of \cite{gv} to apply, however, it was important that the number of D2 branes
bound to D0 branes is independent of the number of D0s. In our case, we would have a bound state problem involving D4, D2 and D0,
and it is not clear that the number of bound states is independent of the number of D2s along the direction of the
reduction vector $z$. On the other hand, we expect the
number of BPS states to depend only on the norm of the vector of charges in the cohomology lattice of the fiber (\ref{pnorm}), as in the
counting on K3 surfaces considered in \cite{vafa,hmal}. Since the D2 charge
$n_2$ we are summing over in (\ref{gvmod}) is part of the $\Gamma_d^{1,1}$ sublattice, and we are setting $m_2=0$ in the complementary
direction, the norm of the charge vector does not depend on $n_2$, therefore the degeneracies are independent of $n_2$ as in the
situation considered in \cite{gv}.

\begin{table}
\begin{center}
\begin{tabular}{| l | r r r rrr r r r  r|}
\hline
$g$ &$r_u^2/2=-1$ &0 &1 &2       &3 & 4& 5& 6& 7& 8 \\ \hline
0&-2&  0& -24& 0& -180& 0& -1040& 0& -5070 &0\\
1&0 &  4& 12 & 64 & 172 & 576& 1464& 3840& 9396&21056 \\
2&0& 0 & -6& -32& -162& -576& -1980& -5760& -16470 &-42112\\
3&0 & 0 &0  &8  &60  &336  &1420  & 5280 & 17340&52640\\ \hline
\end{tabular}
\caption{BPS invariants $(-1)^{n+m} n_g(r_{\rm u})$ for the untwisted sector, counting
bound states of D0-D2-D4 branes on the Enriques fibre.}\end{center}
\label{bortab}
\end{table}
We then propose that (\ref{bgen}) is a generating functional for an index that counts
BPS particles keeping track of their helicities, as in \cite{gv}. These BPS particles are
obtained from bound states of D4s wrapping the Enriques fiber, D2s wrapped around the curves
in the $E_8$ sublattice of the Enriques cohomology, and D0s. As we argued above,
a formula similar to (\ref{gvhet}) should hold after taking into account the rescaling of the string coupling
constant (\ref{rescaling}) and the fact that we have
two different sectors (untwisted and twisted). The natural labels for D-brane charges in the
untwisted and twisted sector are, respectively,
\be
\label{rvectors}
r_{\rm u}=(n,m, \vec q), \qquad r_{\rm t}=(2m+1, 2n+1, \vec q).
\ee
The difference between the D-brane charges in the two sectors is due to the fact that the D-branes in the untwisted sector are
double covers of the D-branes in the twisted sector. For example \cite{bdm}, the twisted sector
contains 4 D4 branes which have half the charge of a D4 brane in the untwisted sector. These ``fractional" branes
differ in their torsion charge. Bound states in the twisted sector are made out of one of these ``fractional" D4 branes together with an arbitrary number of
D4 branes with integer charge. This is why the D4-brane charge in the twisted sector is of the form $2m+1$.
In order to count the bound states, we introduce two sets of BPS invariants for the twisted and untwisted
sectors, $n_g(r_{\rm u})$ and $n_g(r_{\rm t})$. Our proposal for their generating functionals is the natural one from the above
results:
\be
\begin{aligned}
\sum_{r_{\rm u}} (-1)^{n+m} n_g(r_{\rm u}) z^{g} q^{r_{\rm u}^2/2} =&f_1(q) \xi^2(z,q),\\
\sum_{r_{\rm t}}  (-1)^{n+m}n_g(r_{\rm t}) z^{g} q^{r_{\rm t}^2/2} =&-4f_1(q) \xi^2(z,q^4),
\end{aligned}
\ee
where the norm of the vectors are computed as before in the lattice $\Gamma^{1,1}\oplus E_8(-2)$. Notice that the extra factor
of $4$ in the twisted sector corresponds to the D-brane states that differ in torsion classes. This factor is in turn related to the
four hypermultiplets in the fundamental representation of $SU(2)$ that appear in the $N_f=4$ degeneration of the type IIA theory. We show some values of
these BPS invariants in tables 2 and 3.

\begin{table}
\begin{center}
\begin{tabular}{| l | r r r r r r r |}
\hline
$g$ &$r_t^2/2=-1$ & 1&3 &5       &7 & 9& 11 \\ \hline
0&8&  96& 720& 4160& 20280& 87264& 340912 \\
1&0 &  0& -16& -192 & -1488& -8896& -44944 \\
2&0& 0 & 0& 0&  24&  288&  2288 \\
3&0& 0 & 0& 0&  0&  0&  -32 \\
 \hline
\end{tabular}
\caption{BPS invariants $(-1)^{n+m} n_g(r_{\rm t})$ for the twisted sector. Notice that $r_t^2/2$
is always odd, as follows from (\ref{rvectors}).}\end{center}
\label{borttab}
\end{table}

\sectiono{Geometric computation of the BPS invariants in the fibre}
\label{sec:BPStables}
In this section we will analyze the heterotic predictions for the $F_g$ amplitudes
in the geometric reduction. We will extract the GV invariants and show that they fit
with the geometrical approach developed in \cite{kkv}. This will support our interpretation
of the geometric reduction as the one corresponding to the counting of D2/D0 bound states.

As we already mentioned in section 2, the free energy of the perturbative topological
string can be written in terms of BPS or GV invariants $n_g(r)$. The expansion (\ref{gova})
implies that the BPS invariants of non-constant maps $r\neq 0$ contribute to $F_g$ as
\begin{equation}
F_g=\sum_r {\rm Li}_{3-2g}({\rm e}^{- r\cdot t})\sum_{h=0}^g
a_{h,g} n_h(r) \ ,
\label{fgbps}
\end{equation}
with $a_{g,g}=1,a_{g-1,g}=-(g-2)/12,\ldots,a_{2,g}=2(-1)^g/(2g-2)!,a_{1,g}=0,
a_{0,g}=|B_{2g}|/(2 g (2g-2)!)$.
With (\ref{fgbps}) we rewrite now (\ref{heteroticprediction}) in terms of
the BPS invariants. In the type II picture they are expected to correspond
to an integer index in the cohomology of the moduli space of D2 branes wrapping curves
in the fiber direction of $M$. Let us now extract these invariants from (\ref{heteroticprediction}).

If at least one entry in $r\in \Gamma_E=\Gamma^{1,1} \oplus E_8 (-1)$ is odd, the second term in
(\ref{heteroticprediction}) does not contribute
and we get the invariants $n^{\rm odd}_{g}(r)$ listed in the table 4.
Note that $r^2\in 2 \ZZ$ because $\Gamma_E$ is even. In particular the
prediction that $n^{\rm odd}_{g=0}(r)=0$ follows from the fact that the modular form $f_1(q)$ in (\ref{fhsvforms})
has no even powers.
The fact that all $n^{\rm odd}_{g}(r)$ are integers is an important check on
the consistency of the calculation.

\begin{table}
\label{oddbpstab}
\begin{center}
\begin{tabular}{|l | r r r r r r r r r |}
\hline
$g$ &$r^2=0$ &2      &4       &6 & 8& 10& 12& 14& 16 \\  \hline
0&0&  0& 0& 0& 0& 0& 0& 0& 0\\
1&8&  128& 1152& 7680& 42112& 200448& 855552& 3345408& 12166272\\
2&0&  -16& -288& -2880& -21056& -125280& -641664& -2927232& -12166272 \\
3&0&  0& 24& 480& 5264& 41760& 267360& 1463616& 7096992\\
4&0& 0& 0& -32& -704& -8400& -71872& -492800& -2872512\\
5&0&  0& 0& 0& 40& 960& 12384& 113728& 831960\\
6&0&  0& 0& 0& 0& -48& -1248& -17312& -169920\\
7&0&  0& 0& 0& 0& 0& 56& 1568& 23280\\
8&0&  0& 0& 0& 0& 0& 0& -64& -1920\\
9&0&  0& 0& 0& 0& 0& 0& 0& 72\\  \hline
\end{tabular}
\caption{BPS invariants $n^{\rm odd}_{g}(r)\in \ZZ$
for the odd classes $r$ in the fiber direction.}\end{center}
\end{table}

If all entries in $r$ are even then $r^2\in 8 \ZZ$ and
we call the class $r$ even. In (\ref{heteroticprediction})
the second  term gives a subleading correction to
$n^{\rm even}_g(r)$, and we again find that all of them are integer. The first few are listed in table 5.

\begin{table}
\begin{center}
\begin{tabular}{| l | r  r r  r r |}
\hline
$g$ &${r^2}=0$&  8          &16      &24       &32 \\\hline
0&0& 0& 0& 0& 0 \\
1&4& 42048& 12165696& 1242726144& 69636018752 \\
2&0& -21024& -12165696& -1864089216& -139272037504 \\
3&0& 5256& 7096656& 1708748448& 174090046880 \\
4&0& -704& -2872416& -1158884992& -165915421248\\
5& 0& 40& 831948& 611668944& 127601309256\\
6& 0& 0& -169920& -254819136& -80867605120\\
7& 0& 0& 23280& 83673040& 42545564896 \\
8& 0& 0& -1920& -21406464& -18592299200 \\
9& 0& 0& 72& 4174920& 6721882484  \\
10& 0& 0& 0& -598848& -1994908928\\
11& 0& 0& 0& 59472& 480175264\\
12& 0& 0& 0& -3648& -92117568\\
13&0& 0& 0& 104& 13732280 \\
14&0& 0& 0& 0& -1531072\\ \hline
\end{tabular}
\label{evenbpstab}
\caption{BPS invariants $n^{\rm even}_{g}(r)\in \ZZ$ for the even classes $r$ in the fiber direction.}
\end{center}
\end{table}

In \cite{gv} the numbers $n_g(r)$ were given a geometrical interpretation.
In simple cases they can be computed as an index on the cohomology $H^*({\cal M})$
of the moduli space ${\cal M}$ of D2 branes \cite{kkv}.
By ``simple" we mean that the D2 wraps an irreducible and possibly mildly nodal
curve $C\in M$ in the class $r$.  Infinitesimally, the moduli space is
parameterized by the zero modes on
the D2 brane. These form a supersymmetric spectrum with $2g$ zero modes
of the flat $U(1)$ connection on $C$ parameterizing the Jacobian ${\rm Jac}(C)\sim \bT^{2g}$.
Furthermore, there are $h^0({\cal O}(C))$ zero modes corresponding
to the deformations ${\cal M}_C$ of the curve $C$. For this reason the total
moduli space ${\cal M}$ is expected to have a fibration structure
\be
\begin{array}{ccc}
{\rm Jac}(C)& \longrightarrow & {\cal M} \\
 & & \downarrow\\
 & & {\cal M}_C.
 \end{array}
\ee
The cohomology $H^*({\cal M})$ has a natural  ${\rm su}(2)_L\times {\rm su}(2)_R$
Lefshetz action which corresponds to the spacetime helicities of BPS bound states.
The ${\rm su}(2)_R$ is essentially generated by the K\"ahler
form of the base and the ${\rm su}(2)_L$ by the one of the fibre. The dimension of the cohomology group with
eigenvalues ${j^3_L,j^3_R}$, $N_{j^3_L,j^3_R}(r)$, is not invariant under complex structure
deformations. However, the index  $n_g(r)$ defined in (\ref{index}) is an invariant. In general, it is not clear how to define the
Lefshetz actions on ${\cal M}$. However, in \cite{kkv} the problem was bypassed
by using the Abel-Jacobi map, and the following
formula for the $n_g(r)$ was derived
\begin{equation}
n_{g-\delta}(r)=(-1)^{({\rm dim}({\cal M}_C)+\delta)}\sum_{p=0}^\delta
b_{g-p,\delta-p} \chi({\cal C}^{(p)}), \quad b_{g,k}:=
{2\over k!} \prod_{i=0}^{k-1}(2 g-(k+1)+i), \quad b_{g,0}:=1 \ .
\label{eq:kkvformula}
\end{equation}
Here, ${\cal C}^{(p)}$ is the moduli space of the curve $C$ in the class $r$
together with a choice of $p$ points, which correspond to nodes of $C$.
In particular ${\cal C}^{(0)}={\cal M}_C$.
In (\ref{eq:kkvformula}) $\delta$ is  the number of nodes and the formula
is applied as follows. In the simplest situation $C$ is a smooth curve
of genus $g$ in the class $r$, then $\delta=0$ and
\be
n_{g}(r)=(-1)^{{\rm dim}({\cal M}_C)} \chi({\cal M}_C).
\ee
This can be
understood directly as follows. If $C$ is smooth the ${\rm Jac}(C)$
is non-degenerate and carries $I_g$ as ${\rm su}(2)_L$ Lefshetz
representation of the fibre. The sum over $j_R^3$ in
(\ref{index}) gives --up to sign--- the Euler number of the base.
If the contribution to $n_{g-\delta}(r)$ comes only from an
irreducible curve with $\delta$ nodes we can calculate in certain
situations $\chi({\cal C}^{(p)})$ to obtain the BPS number.

We now apply these ideas to D2 branes wrapping curves $C$ in
the fibre of the Calabi-Yau manifold $M$ of the FHSV model.
The moduli space ${\cal M}_C$ factorizes for these curves
into ${\cal M}_C(F)$, parameterizing movements of $C$ in the
fibre, and $\IP^1$, parameterizing movements of $C$ over the base
of $M$. Along this $\IP^1$ direction and outside the $p_i$, the
${\rm Jac}(C)$ is constant. The $\IP^1$ is therefore a component, whose
contribution factors in (\ref{index}). Moreover on this component
the ${\rm su}(2)_R$ Lefshetz action in (\ref{index}) reduces its
contribution to an integral over the Euler class $\int_{\IP^1} e$. This integral localizes to the $p_i$.
The relevant part of the D2 brane moduli space to curves $C$ in the fiber
hence localizes to curves which sit in
the Enriques fibre.

It is  therefore sufficient to consider curves in the four special
Enriques fibres to explain the BPS numbers in the tables in
Sec. \ref{sec:BPStables}. Let us first recall an important fact
about curves in an Enriques surfaces.  According to
proposition 16.1 in \cite{bpv}, for every such $C$ in the class $r$
in the K\"ahler cone there is a second curve $C+K_E$ in
the class $r$ up to torsion with $|C+K_E|\neq \emptyset$
and $r^2=[C]^2=[C+K_E]^2$. So each curve in the Enriques fibre
is effectively doubled. Since we have four fibers we expect
that the numbers in tables 4 and 5 are divisible by eight, which
is indeed the case. Let us now compute the moduli space of
deformations ${\cal M}_C$ for smooth curves of genus $g$.
By the adjunction formula, for a curve $C$ in the class $r=[C]$ we have
\be
2g-2=[C]^2+[C][K_E]\ ,
\label{eq:adjunction}
\ee
where the second term $[C][K_E]=0$ on an Enriques surface.
The moduli space of the curve $C$ in a surface $S$ is given by
the projectivization of ${\cal M}_C=\IP H^0({\cal O}(C))$, and the
dimension $h^0({\cal  O} (C))$ can be calculated  using the
Riemann-Roch theorem
\be
\chi({\cal O}(C))={[C]^2+[C][K_S]\over 2}+\chi({\cal O}(S)),
\label{eq:riemannroch}
\ee
where $\chi({\cal O}(C))=\sum_{i=1}^n (-1)^i h^i({\cal O}(C))$. For smooth
curves in an Enriques surface, $H^1({\cal O}(C))=H^2({\cal O}(C))=0$
and  $\chi({\cal O}(C))=h^0({\cal O}(C))$. Moreover  from section 3.2
we know that $\chi({\cal O}(E))=\sum_{i=1}^2 h^{i,0}(E)=1$, and combining that
with (\ref{eq:adjunction},\ref{eq:riemannroch})  yields
\be
{\cal M}_C=\IP^{g-1}\ .
\label{eq:smoothfibre}
\ee
We apply now (\ref{eq:kkvformula}) and get, for smooth curves in the class $r$
of genus $g={r^2\over 2}+1$,
\be
n_{g}(r)=8\cdot (-1)^{{r^2\over 2}} \chi(\IP^{r^2\over 2})=8 \cdot (-1)^{r^2\over 2} \Bigl({r^2\over 2}+1\Bigr)\
\label{eq:smooth}
\ee
in agreement with table 4.

Let us now give a more detailed calculation involving
the nodal curves. The task is to calculate the Euler numbers $\chi({\cal C}^{(\delta)})$.
If we force the smooth curve $C$ to pass through $\delta$ given
points in $E$, corresponding to the locations of the nodes, we impose $\delta$
linear constraints on its moduli space ${\cal M}=\IP^{g-1}$. The
moduli space of deformations is therefore reduced to
$\IP^{g-\delta-1}$. On the other hand we are free to choose
the position of the points, which are therefore
part of the moduli space of the nodal curves. The freedom of
choosing $n$-points on $E$ is naively $E^n$. Since the
points are undistinguishable one considers the orbifold
${\rm Sym}^{n}(E)=E^n/S_n$ by the symmetric group $S_n$.
 The relevant model for the moduli space of $n$ points is the
``free field'' resolution \cite{yz} ${\cal M}_\delta={\rm Hilb}^\delta(E)$
of this orbifold. The name comes from the fact that the Euler numbers of the resolved
spaces are generated by a free field representation
\be
\sum_{n=0}^\infty \chi({\cal M}_n) q^n=\prod_{n=1}^\infty \left(1\over 1-q^n\right)^{\chi(E)}=
1+12 q+90 q^2 +  520 q^3 + 2535 q^4 +\ldots
\ee
This is special bosonic case of a formula \cite{gs}, which gives the Poincar\'e polynomial
of ${\cal M}_n$ in terms of bosonic and fermionic free fields. The reason that one needs
only 12 bosons here is that $E$ has only even cohomology. Since ${\rm Hilb}^\delta(E)$
fibers trivially over $\IP^{g-\delta-1}$
we obtain
\be
\chi({\cal C}^{(\delta)})=(g-\delta)\chi({\cal M}_\delta).
\label{eq:ecd}
\ee
If we insert this result into (\ref{eq:kkvformula}) we reproduce immediately,
and to a large extent, the heterotic predictions in table 4. The deviations between the
two calculations are given in table 6.  As we will see later, the heterotic predictions are in full
agreement with the computation of the topological string amplitudes by using the B model, and
the deviations recorded in table 6 are due to the fact that for reducible curves with many nodes one has to
refine the computation of BPS invariants in (\ref{eq:kkvformula}) as explained in \cite{kkv}.
\begin{table}
\begin{center}
\begin{tabular}{| l | r r r rrr r r r |}
\hline
$g$ &$r^2=0$ &2      &4       &6 & 8& 10& 12& 14& 16 \\ \hline
0&0&  0& 0& 0& 0& 0& 0& 0& 0\\
1&0 &  0 & 0 & 0 & 24& 288& 2160& 12544& 61608 \\
2&0& 0 & 0& 0& 0& 0& 0& -32& -384 \\
3&0 & 0 &0  &0  &0  &0  &0  & 0 & 0\\ \hline
\end{tabular}
\caption{Differences between the heterotic BPS
prediction in table 2 and the geometric BPS calculation using
(\ref{eq:kkvformula},\ref{eq:ecd}).}\end{center}
\label{devtab}
\end{table}

It is instructive to compare this situation with the K3 curve counting of \cite{yz}.
Because of $\chi({\cal O}(K3))=2$ we get in that case ${\cal M}_C= {\IP}^{g}$, instead of
(\ref{eq:smoothfibre}). On K3 we can force up to
$g$ nodes to get rational curves. For the Enriques surface, the smooth moduli
of a genus $g$ curve is too small to allow generically for $g$ nodes.
This also explains the absence of genus zero invariants.

We close this section by noticing that the genus one BPS invariants for odd classes listed in table 4 
have an 
interesting algebraic interpretation. As we pointed out after (\ref{anotherborcherds}), 
$F_1(t)$ can be interpreted as the logarithm of a denominator formula of a superalgebra. The 
genus one BPS invariants $c_1(r^2)$ are then multiplicities of the (super)root spaces of this 
superalgebra. 

\sectiono{The B-model for an algebraic realization of the FHSV model}

In this section we find an algebraic realization for the double cover of the
Enriques CY, and we study it and its $\ZZ_2$ quotient using mirror symmetry.
To simplify the analysis we define a ``reduced" FHSV model by blowing down
an $E_8$ part of the Picard lattice. The reduced model has only three K\"ahler moduli,
and its mirror can be analyzed in detail in the context of the B-model. We show that
the topological string of the reduced model can be solved in terms of modular forms
through the arithmetic properties of the mirror map. Using the holomorphic anomaly equations of \cite{bcov} we find
explicit, closed expressions for the topological string amplitudes up to
genus $4$. In this section we will denote genus $g$ amplitudes by $F^{(g)}$.

\subsection{The geometric description of the Enriques Calabi-Yau}

Here we describe the periods of the Enriques Calabi-Yau and we find
an algebraic description of the double cover and its mirror.

\subsubsection{Periods and prepotential}
\label{sec:periodsandprepotentential}
If $M$ is a $d=$dim$_\IC$ dimensional CY manifold and
$\omega_{d,0}$ is its unique holomorphic $(d,0)$-form,
we may consider the quantities
\be
W^{(\bf k)}=\int_{M} \omega_{d,0}
\partial^{\bf k} \omega_{d,0},
\ee
where $\partial^{\bf k}=
\partial^{k_1}_{z_1}\ldots\partial^{k_{\rho-2}}_{z_{\rho-2}}$ denotes
derivatives w.r.t. to the complex moduli. If  the order of the
derivative operator is $\sum k_i=|{\bf k}|$ then Griffiths
transversality \cite{BryantGriffith} implies
\be
W^{(\bf k)}=0\ , \quad {\rm if}  \ |{\bf k}|<d \ .
\label{eq:griffith}
\ee
In particular, for $d\in 2 \ZZ$ we get from $W^{(0)}=0$
an algebraic relation between the periods, while for $d=3$ eqs.
(\ref{eq:griffith}) lead to ${\cal N}=2$ special geometry.

Let us describe first properties of the periods of the manifold
$M=({\rm K3}\times \bT^2)/\ZZ_2$, which follow from the
double cover. On the K3 covering of the Enriques surface
we can choose a twelve-dimensional basis of two-forms, $\alpha_i$, $i=0,\ldots,\rho-1$, in
the anti-invariant lattice $\Gamma_{\rm K3}^-$ and satisfying
\be
\int_{\rm K3} \alpha_i \wedge \alpha_j=\eta_{ij}.
\ee
Here $\eta_{ij}$ is the symmetric, even intersection form on $\Gamma^-_{K3}$.
We consider families of K3 surfaces covering the Enriques surface where
$\rho=12$ is the number of anti-invariant transcendental cycles, i.e. we
choose a polarization so that the dual cycles, with basis
$\Gamma^j$ such that $\int_{\Gamma^j} \alpha_i=\delta_i^j$,
are transcendental. We have  $\Gamma^i\cap \Gamma^j=\eta^{ij}$
with $\eta_{ij} \eta^{jk}=\delta_i^k$. The discussion below
holds for general algebraic K3 surfaces with $\rho$ transcendental cycles,
and in particular for the reduced model
for which $\rho=4$.

The holomorphic $(2,0)$ form can be expanded as
\be
\omega_{2,0}=\sum_{i=0}^{\rho-1}
\hat X^i\alpha_i\ \qquad {\rm with } \quad \hat X^i=\int_{\Gamma^i} \omega_{2,0} .
\ee
The period integrals $\hat X^i(z)$ depend on the complex structure
deformation parameters $z_a$, $a=1,\ldots,\rho-2$, that appear
in the algebraic definition of the model. Griffiths
transversality (\ref{eq:griffith}) implies for $|{\bf k}|=0$
\begin{equation}
\sum_{i,j=0}^{\rho-1}\hat X^i \hat X^j \eta_{ij}=0.
\label{periodrelation}
\end{equation}
and for $|{\bf k}|=1$
\be
\begin{aligned}
{3\over 2} \partial_1 W^{(2,0,0)}=W^{(3,0,0)},\quad
{1\over 2} \partial_2 W^{(2,0,0)}+\partial_1 W^{(1,1,0)}&=W^{(2,1,0)},\\
{1\over 2} (\partial_1 W^{(0,1,1)}+\partial_2 W^{(1,0,1)}+\partial_3 W^{(1,1,0)})&=W^{(1,1,1)}\ .
\end{aligned}
\ee
These can be integrated using the Picard-Fuchs (PF) equations
and  yield rational expressions for the B-model
two-point function $W^{(\bf k)}$, $|{\bf k}|= 2$ in terms
of the complex structure deformation parameters $z_k$. We will denote them as
$C_{z_i,z_j}:=W^{(\bf k)}$, $|{\bf k}|= 2$, where ${\bf k}$ has an
entry $2$ at the $i$'th position if $i=j$ and entries $1$
at the $i$'th and $j$'th position otherwise.

There is a maximal unipotent point in the
moduli space of complex structures of the K3 at which one has
one unique holomorphic solution ${\hat X}^0$,
$\rho-2$ single logarithmic solutions ${\hat X}^a$,
$a=1,\ldots,\rho-2$ and one double logarithmic
solution $\hat F:=\hat X^{\rho-1}$. We define the mirror map
for the K3 as
\begin{equation}
\hat t^a(z)={\hat X^a\over \hat X^0}(z)\qquad a=1,\ldots,\rho-2 .
\label{mirrormapK3}
\end{equation}
The couplings $C_{z_a,z_b}$ transform like sections of the bundle
${\rm Sym}^2 T^*{\cal M}\otimes {\cal L}^{-2}$,
where ${\cal M}$ is the moduli space of complex structures on the K3 and ${\cal L}$ is the
K\"ahler line bundle.
The mirror map and the special gauge of the B-model w.r.t. to the K\"ahler line bundle
relates then the B-model couplings to their  A-model counterparts in the following way
\be
C_{t^a t^b}=
{1\over ({\hat X^0}(z(t)))^2} C_{z_c z_d}(z(t))
{\partial z_c\over \partial t^a}(z(t)){\partial z_d\over \partial t^b}(z(t))=\hat \eta_{ab},
\label{two-point}
\ee
where $\hat \eta_{a b}$ are the classical intersection numbers
of the generators of the Picard lattice of the mirror K3, related
to the $\eta_{ij}$ above by
\be
\eta_{ij}=h_{0,\rho-1}\oplus \hat \eta_{ab},\qquad
h_{0,\rho-1}=\left(\begin{array}{cc} 0& 1 \\ 1 & 0\end{array} \right).
\ee
Eq. (\ref{two-point}) is a consequence of (\ref{eq:griffith}) and reflects the
absence of instanton corrections to the classical intersection ring of K3. It
yields a simple relation between the quantity $\hat X^0(z(t))$, which corresponds to a gauge choice
of $\omega_{2,0}$ in ${\cal L}^{-1}$, and the derivatives
${\partial z_i\over \partial t_a}(z(t))$ of the mirror map. The latter
is a total invariant under the subgroup $\Gamma_{X}$ of the discrete
automorphism group ${\rm Aut}(\Gamma^-_{K3})$, that is realized as monodromy
on the periods $(\hat X^0,\ldots,\hat X^{\rho-1})$ on the algebraic family $X$ double
covering the Enriques surface $E$. The ability to express $\hat X^0(z(t))$ using (\ref{two-point}) as
a modular form of $\Gamma_X$ will become important to solve for the
$F^{(g)}$ in sec. \ref{sec:bmodelrestricted}.
We can write the periods as a vector of inhomogeneous coordinates
\be
\hat \Pi=
\left(\begin{array}{c} \hat X^0\\ \hat X^1\\ \vdots\\ \hat X^{\rho-2}\\ \hat
X^{\rho-1}\end{array}\right)=
\hat X^0\left(\begin{array}{c} 1 \\ t^1\\ \vdots \\ t^{\rho-2}\\ -{1\over 2}\hat
\eta_{a,b} t^a t^b- 1\end{array}\right)\ ,
\ee
where we used the mirror map (\ref{mirrormapK3}) as well as
(\ref{periodrelation}) and the explicit form of $\eta_{ij}$. Note that
the $-1$ in the last period is a specialization to $d=2$ of the
${X^0\over (2 \pi i)^d} \zeta(d)\chi(M)$ term, which appears
in $d=3,4$ CY manifolds.

Similarly we define for the $\bT^2$ a basis of anti-invariant one
forms $\alpha,\beta$ with $\int_{\bT^2}\alpha \wedge \beta=1$ and all other
integrals are zero. We expand
\be
\omega_{1,0}=x^0\alpha-x^1\beta
\ee
where the coefficients $x^0$, $x^1$ are the periods
\be
x^0=\int_a\omega_{1,0}, \qquad x^1=\int_b\omega_{1,0}
\ee
and $\int_a \alpha=-\int_b \beta=1$. At the point of
maximal unipotent monodromy for the $\bT^2$ we have a regular solution for
$x^0$ and a logarithmic solution for $x^1$, and we define the mirror map
as
\be
\tau(z)={x^1\over x^0}(z).
\label{mirrormapT2}
\ee
With similar definitions as
above we get a one-point function from integrating
$\partial_1 W^{(1)}=W^{(2)}$. The analogue of (\ref{two-point})
yields well known relations between the $j$-function and
the Schwarz triangle functions for subgroups $\Gamma_{\bT^2}$
of $SL(2,\ZZ)$ (see, for example, \cite{kly}).

We combine information about $\bT^2$ and K3 to write periods of $M$.
The 3-cycles
of $M$ are
\be
\begin{array}{rlrl}
A_0&=a\times \Gamma_0, & B_0&=b\times \Gamma_{\rho-1},\\
A_i&=a\times \Gamma_i, & B_i&=b\times \sum_{j=1}^{\rho-2}\eta_{ij} \Gamma_j,
\qquad i=1,\ldots \rho-2,\\
A_\tau&=b\times \Gamma_0, & B_\tau&=a\times \Gamma_{\rho-1}.
\end{array}
\ee
This basis is symplectic with $A_i\cap B_j=\delta_{ij}$.
The invariant holomorphic $(3,0)$-form of $M$ $\Omega$ is given
by $\Omega=\omega_{2,0}\omega_{1,0}$ which in the
algebraic model is realized as
\be
\Omega= {\dd x \over  y } \wedge {s u {\dd t
\wedge \dd v\over {\cal Y}}}.
\ee
If we integrate this e.g. over
$A_0$ we get
\be
X^0=\int_{A_0}\Omega=\int_{a}
\left( \dd x \over  y \right)\int_{\Gamma_0}\left(s u {\dd t \wedge \dd
v\over {\cal Y}}\right)=x^0 \hat X^0 \ .
\label{X0}
\ee
We can write the period vector
of the threefold in a symplectic basis
\be
\Pi=
\left(\begin{array}{c}
\int_{B_0}\Omega \\ \vdots \\\int_{B_{\tau}}\Omega \\
\int_{A_0}\Omega \\ \vdots \\\int_{A_{\tau}}\Omega \\
\end{array}\right)=
\left(\begin{array}{c} - x^1  \hat X^{11}\\
-x^1 \eta_{1 a} \hat X^a\\ \vdots\\ x^0 \hat X^{11} \\ x^0 \hat X^0 \\
x^0 \hat X^1 \\ \vdots \\ x^1 \hat X^0\end{array}\right)=
X^0\left(\begin{array}{c} 2 {\cal F} - t ^i \partial_i {\cal F}-\tau \partial_\tau {\cal F}\\
\partial_1 {\cal F}\\ \vdots\\ \partial_{\tau} {\cal F} \\ 1 \\
t^1\\ \vdots \\ \tau\end{array}\right)\ ,
\ee
{}From the above we read off the prepotential
\be
F_0=- {1\over 2} \tau \hat \eta_{a,b} t^a t^b - \tau
\ee
and conclude that there are no
instanton corrections at genus $0$ in base, fibre and mixed directions.

$\Gamma_{\bT^2}$ acts on the periods $(x^0,x^1)$ a subgroup of $SL(2,\ZZ)$.
Similarly, $\Gamma_X$ acts on $(\hat X^0,\ldots,\hat X^{\rho-1})$. The action of these
two groups on $\Pi$ does not commute and generates a bigger discrete group
$\Gamma_M$. From the point of view of the heterotic string dual, $\Gamma_X$ generates $T$-dualities,
$\Gamma_{\bT^2}$ generates $S$-dualities, and $\Gamma_M$ is
called the U-duality group.

\subsubsection{Mirror symmetry on an algebraic realization}

In order to calculate the periods discussed in the last section we need an
algebraic realization and understand mirror symmetry on it. Let us first explain
some features of mirror symmetry for K3 surfaces. Mirror
pairs of K3 can be given by three dimensional reflexive polyhedra
following Batyrev's mirror symmetry construction \cite{Batyrev}.
Fortunately the double covering of Enriques that \cite{horikawa} uses is
of this type. The small polyhedron $\Delta$ is given by the convex
linear hull of the corners
\be
\{\nu_1,\ldots, \nu_5\}=\{[1,0,1],[0,1,1],[-1,0,1],[0,-1,1],[0,0,-1]\}\ .
\ee
Its points are on a lattice $\Lambda\sim \ZZ^3$, which makes it an
integral polyhedron. $\Lambda$ is obviously embedded like $\Lambda\in \Lambda_\IR=N\sim \IR^3$.
The polyhedron is  reflexive, which means that  the dual $\Delta^*:=
\{x\in M^*|\langle x,y\rangle \ge -1, \forall\ y\in \Delta\}$ is an
integral polyhedron in the dual lattice $\Lambda^*\in N^*$.
The $\nu_1,\ldots,\nu_5$ above are corners of the polyhedron, and the corners
$p^*_i$ of the large polyhedron $\Delta^*$ are found by solving
the equations $\langle \nu_i^*,\nu_{j_k}\rangle =-1$, $k=1,\ldots ,3$:
\be
\{\nu_1^*,\ldots, \nu_5^*\}=\{[2,2,1],[2,-2,1],[-2,-2,1],[-2,2,1],[0,0,-1]\}\ .
\ee
To the polyhedra Batyrev associate hypersurfaces
\be
p=\sum_{i=1}^{\# \nu} a_i \prod_{j=1}^{\# \nu^*} x_j^{\langle \nu_i,\nu^*_j\rangle +1}=0,
\qquad
p^*=\sum_{i=1}^{\# \nu^*} b_i \prod_{j=1}^{\# \nu} y_j^{\langle \nu_i,\nu^*_j\rangle +1}=0,
\label{hypersurfaces}
\ee
which describe mirror manifolds $M$ and $W$. For example we get a special
double cover of $\IP^1\times \IP^1$
\be
p=a_1 s^4 u^4+ a_2 s^4 v^4 + a_3 t^4 v^4+ a_4 t^4 u^4 + a_5 y^2+ a_0 y s t u v=0 \ .
\label{restrictedfamily1}
\ee
Note that we take the product only over the corners of $\Delta^*$, to which
we associate the variables $\{x_j\}=\{s,u,t,v,y\}$, while
we include all points that are not inside codimension one faces
in $\Delta$. Four of the $a_i$ are redundant, i.e. they can be set to
say one  by the automorphism of the ambient space $\IP_\Delta$.

Calabi-Yau threefolds correspond to 4d polyhedra and the
cohomology groups of $M$ and $W$ are given by
the  formulas
\begin{equation}
\begin{array}{rl}
h^{21}(M)&=\ds{l(\Delta)-5 - \sum_{{\rm codim} (\theta)=1}
l^*(\theta)+\sum_{{\rm codim}(\theta^*)=2} l^*(\theta^*) l^*(\theta)}, \\ [ 3 mm]
h^{11}(M)&=\ds{l(\Delta^*)-5 - \sum_{{\rm codim} (\theta^*)=1} l^*(\theta^*)+
\sum_{{\rm codim}(\theta^*)=2} l^*(\theta^*) l^*(\theta)} \ ,
\end{array}
\end{equation}
where $\theta$ is a face ($\Delta$ is the top dim face) and $l(\theta)$ means
all points in that face, while $l^*(\theta)$ means interior
points in that face.

For K3 these formulas become
\begin{equation}
\begin{array}{rl}
\rho(M)-2&=\ds{l(\Delta)-4 - \sum_{{\rm codim} (\theta)=1} l^*(\theta) +
\sum_{{\rm codim}(\theta)=2} l^*(\theta) l^*(\theta^*)} \\
h(M)&=\ds{l(\Delta^*)-4 - \sum_{{\rm codim} (\theta^*)=1}l(\theta^*) +
\sum_{{\rm codim}(\theta)=2} l^*(\theta^*) l^*(\theta)}
\end{array}
\end{equation}
and the interpretation is as follows.
$h(M)$ is the rank of the Picard group of the K3 $M$ and
$\rho(M)$ is the number of transcendental cycles of $M$.
For all 3d  reflexive polyhedra one has $h(M)+\rho(M)=h^{11}(K3)+h^{20}(K_3)+h^{02}=22$,
and the different choices are merely different
choices of the polarization. In our case we have
$\rho(M)=4$ and $h(M)=18$. Here as above for the mirror $W$ we just exchange
$\Delta$ with $\Delta^*$.

For the polyhedra $\Delta$, i.e. the manifold $M$,
we define the variables
\be
\label{csvariables}
z_1={a_1 a_3 a_5^2\over a_0^4}, \qquad z_2={a_2 a_4 a_5^2\over a_0^4}.
\ee
In terms of these variables we have the following Picard-Fuchs equations
\begin{equation}
\begin{aligned}
{\cal L}_1&=\theta_1^2- 4 (4 \theta_1 + 4 \theta_2-3) (4 \theta_1+ 4 \theta_2-1) z_1,\\
{\cal L}_2&=\theta_2^2- 4 (4 \theta_1 + 4 \theta_2-3) (4 \theta_1+ 4 \theta_2-1) z_2,
\end{aligned}
\label{pf1}
\end{equation}
where we defined the logarithmic derivative $\theta_i=z_i {\dd \over \dd z_i}$. The
periods are linear solutions and in particular $z_i=0$ is a point of maximal
unipotency. Around this point we have a pure power series and two single logarithmic
solutions, which correspond to geometric periods. In particular the
mirror map is given by
\begin{equation}\label{mirrormap}
\begin{aligned}
z_1&=q_1 - 40q_1^2 + 1324q_1^3 - 64q_1q_2 + 2560q_1^2q_2 + {\cal O}(q^4),\\
z_2&=q_2 - 64q_1q_2 + 2560q_1^2q_2 - 40q_2^2 + 2560q_1q_2^2 +{\cal O}(q^4),
\end{aligned}
\end{equation}
with $q_i=\re^{-t_i}$, where $t_i$, $i=1,2$ are complexified K\"ahler parameters
of the mirror. Indeed, calculation of the intersection numbers in the polyhedron
$\Delta$ reveals that the corresponding Picard lattice is $\Gamma^{1,1}$.
Using Griffiths transversality (\ref{eq:griffith}) and (\ref{pf1}) we can
evaluate the two-point functions
\begin{equation}
\label{k3yukawas}
C_{x_1x_1}={2 \over x_1 \Delta},\qquad C_{x_1x_2}={1-x_1 - x_2 \over x_1 x_2 \Delta},
\qquad C_{x_2x_2}={2 \over x_2 \Delta}\ .
\ee
Here
\be
\label{discriminant}
\Delta=1-2 (x_1+ x_2 + x_1 x_2) +x_1^2 + x_2^2
\end{equation}
is the principal discriminant, and we rescaled the variables
suitably $z_i=x_i/64$. Due to the above mentioned special properties of
the mirror map of K3, we find that $C_{t_1t_1}=C_{t_2t_2}=0$, $C_{t_1t_2}=1$
in agreement with the identification of the Picard lattice $\Gamma^{1,1}$.

So far we have calculated on the mirror of the double covering
of the Enriques surface and we need a justification that
the two parameter family treated above does appear as
a subfamily of the family of K3 admitting the Enriques involution.
We consider now this symmetric splice of the mirror polynomial
in (\ref{hypersurfaces}). Let us  label the coordinates
$\{y_i\}$ again by $(s:t,u:v,y)$. The $\ZZ_2$ involution that we
want to mod out $(s:t,u:v,y)\mapsto (-s:t,-u:v,-y)$ will break
some of the automorphisms of the ambient space. Therefore we can
not follow Batyrev's methods that restricts to the true deformation
parameters, which amounts to dropping the codimension one points,
as used above.

The geometry of the mirror is again a double covering of
$\IP^1\times \IP^1$ branched at the generic degree
$(4,4)$-hypersurface. The following expression keeps only
monomials that are invariant under the $\ZZ_2$:
\be
\begin{array}{rl}
{\cal Y}^2=&
b_1 t^4 v^4 +
b_2 t^4 u^2 v^2 +
b_3 t^4 u^4 +
b_4 u^4 t^2 s^2 +
b_5 s^4 u^4 +
b_6 s^4 u^2 v^2\\
      &
+b_7 v^4 s^4+
b_8 s^4 t^2 s^2+
b_{9} t^3 u^3 s v +
b_{10} s^3 u^3 t v\\
      &+
b_{11} v^3 s^3 t u +
b_{12} t^3 v^3 u s +
b_{13} u^2 v^2 t^2 s^2
\end{array}
\end{equation}
Normally the $SL(2,\ZZ)$ transformations of the two $\IP^1$ eliminate
each three complex parameters. However to be compatible with the
$\ZZ_2$ we can only make a rescaling $(s,t)\mapsto (\mu s, \mu^{-1} t)$ and
similarly for $(u,v)$. The weighted transformation of $y^2\mapsto
{\cal Y}^2 + {\cal Y} f_{2,2}(s,t,u,v)$ has been used to eliminate linear terms in $y$,
and an overall rescaling will eliminate a third $b_i$. In total we
have hence ten invariant $b_i$, which are precisely the deformation
parameters of the Enriques surface.

It is still very tedious to derive a ten parameter PF system. Let us
therefore look at a further
symmetric restriction forced by the symmetry $u\mapsto i u$ and
$s\mapsto i s$ so that the $\ZZ_2\times \ZZ_4$ invariant subslice has the
form
\be
b_0 {\cal Y}^2=
b_1 t^4 v^4 +
b_3 t^4 u^4 +
b_5 s^4 u^4 +
b_7 v^4 s^4+ b_{13} u^2 v^2 t^2 s^2\ .
\label{restrictedfamily2}
\end{equation}
We will argue now that this family is, up to a $\ZZ_2$ symmetry acting on their
moduli space, identical to the family (\ref{restrictedfamily1}). At first
glance this seems strange, because every monomial in the family
(\ref{restrictedfamily2}) is invariant under  the Enriques involution
$(s:t,u:v,y)\mapsto (-s:t,-u:v,-y)$, while the monomial $ystuv$ is projected
out from (\ref{restrictedfamily1}). We can keep this monomial by considering
an induced action on the moduli space $a_0\rightarrow -a_0$. Now recall, e.g.
from the Landau-Ginzburg description of the CY manifold, that terms
${\partial p \over \partial x_i}$ are trivial in the sense that they do
not change the residua or periods in the compact case. We can hence use
${\partial p\over \partial y}=2 a_5 y - a_0 stuv \sim 0$ and substitute
$y={a_0\over 2 a_5}stuv$ is the last term of (\ref{restrictedfamily1}),
which leads to the parameter identification $b_{13}=\left(a_0^2\over 2 a_5\right)$,
establishing the equivalence of the two families.

\subsubsection{Arithmetic expressions for the B-model quantities}
As discussed in sec. \ref{sec:periodsandprepotentential} we expect to be
able to give arithmetic expressions for the mirror map and the
fundamental period, similar to what is obtained in \cite{klm,ly}.
If we restrict the PF system (\ref{pf1}) to one variable
by setting $z_2=0$ or $z_1=0$ we find the PF operator
\be
\label{ellpf}
{\cal L}=\theta^2- 4 (4 \theta-3)(4 \theta-1)z,
\ee
which corresponds to the $\Gamma(2)$ elliptic
curve
\be
\label{elcurve}
x_1^2=x_2^4+x_3^4+z^{-{1\over 4}}x_1x_2x_3.
\ee
After transforming this curve to the Weierstrass form
we calculate its $j$-function as
\be
\label{jz}
j(q)=1728 J(q)={(1+192 z)^3\over z (1-64 z)^2}.
\ee
Let us now define
\be
\begin{aligned}
K_2&  =\vartheta_3^4+\vartheta_4^4,\\
K_4&  =\vartheta_2^8\ .
\label{basismodularforms}
\end{aligned}
\ee
In terms of these modular forms, the relation (\ref{jz}) can be
inverted to obtain the Hauptmodul of $\Gamma(2)$ as
\be
z(q)={K_4(q)\over 64 K_2^2(q)},
\label{hmg2}
\ee
which is the arithmetic expression of the mirror map for the curve (\ref{elcurve}). The triviality of the one-point coupling
for the elliptic curve
\be
1={1\over \omega_0^2}{\dd z\over \dd t}{1\over z(1-64 z)},
\ee
which is similar to the triviality of the two-couplings in the K3 case (\ref{two-point}),
leads to the following equation for the fundamental period $\omega_0$ of the PF equation (\ref{ellpf})
\begin{equation}
\label{ellfund}
\omega_0^2(q)={1\over 2} K_2(q)\ ,
\end{equation}
where we have used (\ref{hmg2}).

The full PF system (\ref{pf1}) can also be solved arithmetically in terms of (\ref{hmg2})
by using the techniques of \cite{ly}. It can be easily shown that the simple ansatz
\begin{equation}
\begin{aligned}
z_1(q_1,q_2)&=z(q_1) (1- 64 \, z(q_2))=\ds{{K_4\over 64 K_2^2}\left(1- {\tilde K_4\over \tilde K_2^2}\right)}\\
z_2(q_1,q_2)&=z(q_2) (1- 64 \, z(q_1))=\ds{{\tilde K_4\over 64 \tilde K_2^2}\left(1- {K_4\over K_2^2}\right)},
\end{aligned}
\label{analyticmirrormap}
\end{equation}
where we defined $K_2=K_2(q_1),K_4=K_4(q_1)$ and $\tilde K_2=K_2(q_2),\tilde K_4=K_4(q_2)$,
provides an analytic expression for the mirror map (\ref{mirrormap}). One can also find an analytic
expression for the fundamental period of the system (\ref{pf1}) using (\ref{ellfund})
\begin{equation}
\label{anxo}
(\hat X^0)^2(q_1,q_2)=\omega_0^2(q_1) \omega_0^2(q_2)= {1\over 4} K_2 \tilde K_2.
\end{equation}
It is now easy to show that the discriminant (\ref{discriminant}) can be written as
\be
\Delta=(1-64\, z(q_1) - 64 \, z(q_2))^2,
\ee
where $z(q_i)$ is the mirror map (\ref{hmg2}).

In order to analyze the dependence of the model on the $\bT^2$ in the base, it is
convenient to realize it also algebraically, e.g. as degree 6 curve
\be
x_1^6 +x_2^3 + x_3^2 +z^{-1/6} x_1 x_2 x_3=0
\label{basecurve}
\ee
in $\IP(1,2,3)$, which also can be solved arithmetically. The mirror map is determined by the
equation
\be
J(q_3) ={1\over z(q_3) (1-432 z(q_3))},
\ee
which can be explicitly inverted to \cite{kly,klm}
\be
\label{basemm}
z_3(q_3)={2 \over J(q_3) + {\sqrt {J(q_3) (J(q_3)-1728)}}}.
\ee
Finally, the fundamental period is
\be
\label{basefund}
x^0 (q_3)=E_4^{1/4}(q_3).
\ee
Therefore, the fundamental periods $x^0,\hat X^0$ as well as
the mirror map for the reduced model can be expressed as modular forms in the parameters
$t_1,t_2,t_3$  or functions of subgroups of $SL(2,\ZZ)^3$, and
this fact will become very useful in solving the B-model.

It is instructive to compare the reduced model with the Enriques Calabi-Yau
constructed as an orbifold of $(\bT_2)^3$ by $\ZZ^K_2\times \ZZ^E_2$, which acts of
the coordinates of the three tori $(z_1,z_2,z_3)$ as a Kummer involution
and a free Enriques involution
\begin{equation}
\begin{array}{rlll}
K:& -     \phantom{ \left(1\over 2\right)} \ \ ,& \ - \phantom{ \left(1\over 2\right)} \ \   , &\  + \\
E:& +\left(1\over 2\right)\ ,&\   -\left(1\over 2\right)\ , &\   - \\
KE:& -\left(1\over 2\right)\ ,&\  +\left(1\over 2\right)\ , &\    - \ .\\
\end{array}
\end{equation}
Here $-$ indicates a sign change and $\left(1\over 2\right)$ a shift of the
coordinate $z_i$. This model has three  moduli from the invariant sector
of the orbifold, which also have a natural $SL(2,\ZZ)^3$ acting on them.
However the geometry is very different. The rank 18 Picard lattice of the mirror
of the $K3$ family $X$ has
intersection $E_8(-1)\oplus E_8(-1)\oplus H(1)$. As we will verify in detail in
Sec. \ref{sec:bmodelrestricted}, the reduced model is obtained by contracting the curves in the $E_8(-1)$ part
of the Picard lattice, after the
Enriques identification. If we denote their complex volumes by $t_i$, $i=1, \cdots, 8$,
this is achieved by setting $t_i=0$, i.e. $q_i=\re^{-t_i}=1$.
On the other hand the rank 18 Picard lattice of the Kummer K3 $K$
that emerges after the $\ZZ_2^K$ orbifold in the first two
coordinates is generated over $\IQ$ by sixteen $\IP^1$'s
that resolve the $A_1$ singularities at $P_{ij}$, $i=1,\ldots,4$,
$j=1,\ldots,4$, and two invariant classes from the $\bT_4$. If we
call the resolution map $\sigma:K\rightarrow \bT_4/\ZZ_2^K$ and the exceptional divisors
$E_{ij}:=\sigma^{-1}(P_{ij})$ we get $E_{ij} \cap E_{kl}=-2 \delta_{ik}\delta_{jl}$.
Obviously the lattices that get contracted to reach the orbifold point
and the reduced model are quite different. Moreover due to
the non-trivial $B$-field one expects that the complex volumes of
the $\IP^1$'s $\tilde t_i$ approach $|\tilde t_i|\rightarrow
\pi {\rm i}$ at the orbifold limit so that $\tilde q_i=-1$.
This might explain why all information about the full lattice
disappears in a type II one-loop computation of $F_1$ in the invariant
sector of the orbifold \cite{bcov1}.

\subsection{Topological string amplitudes from the reduced B-model}
\label{sec:bmodelrestricted}

In this subsection we use the holomorphic anomaly equations of \cite{bcov} to
compute the topological string amplitudes for the reduced model in the fiber directions,
up to (and including) $g=4$.

\subsubsection{Genus one amplitude}
As explained in \cite{bcov} the topological or holomorphic limit
of the genus one free energy $F^{(1)}$ is related to the
holomorphic Ray-Singer torsion \cite{raysinger}. The latter
describes aspects of the spectrum of the Laplacians of
$\Delta_{V,q}=\bar \partial_V \bar \partial_V^\dagger+ \bar \partial_V^\dagger \bar \partial_V$ of
a del-bar operator $\bar \partial_V: \wedge^q \bar T^*\otimes V
\rightarrow \wedge^{q+1} \bar T^*\otimes V$ coupled to a
holomorphic vector bundle $V$ over $M$. More precisely with a
regularized determinant over the non-zero mode spectrum  of $\Delta_{V,q}$
one defines\footnote{\cite{Pestun:2005rp} reviews these facts and relates
the Ray-Singer torsion to Hitchin's generalized 3-form action at one loop.}\cite{raysinger}
\be
I^{RS}(V)=\prod_{q=0}^n \bigl({\det}^{ \prime} \Delta_{V,q}\bigr)^{{q\over 2} (-1)^{q+1}}.
\ee
One case of interest, $V=\wedge^p T^*$ with
$\Delta_{p,q}:=\Delta_{\wedge^p T^*,q}$, leads to the definition of
a family index
\be
F^{(1)}={1\over 2}\log \prod_{p=0}^n \prod_{q=0}^n
\bigl( {\det}^{\prime}  \Delta_{pq}\bigr)^{(-1)^{p+q} pq}
\ee
depending only on  the
complex structure of $M$. As was shown in \cite{bcov} the
holomorphic and antiholomorphic dependence of this object
on the complex structure \cite{bgs} can be integrated using
special geometry to yield
\be
F^{(1)}={1\over 2} \log\left[f_1(z)\det \left(\partial z \over \partial t\right)
\over (X^0)^{\kappa} \right].
\label{f1top}
\ee
In this expression, $X^0$ is the fundamental period of the PF system, $\partial z /\partial t$ is
the Jacobian of the mirror map, $\kappa=3+h_{11}-{\chi\over 12}$ depends on global
topological data, and $f_1(z)$ is the holomorphic ambiguity at genus one.
Up to the normalization factor ${1\over 2}$ this is the
same expression that was derived in \cite{bcov1} using
world-sheet arguments. The large volume behavior
\be
F^{(1)}\rightarrow  \sum_{i=1}^{h^{11}}
t_i \int_M c_2(T)\wedge J_i, \qquad t\rightarrow \infty,
\ee
as well as local topological
data of other singular limits in the complex structure moduli space,
determine the leading behavior of $F^{(1)}$ and fix
the holomorphic ambiguity $f(z)$.

We can now use our solution to the variations of Hodge structures for the
family (\ref{restrictedfamily1}) to compute $F ^{(1)}$ for this model.
Indeed our variable choice at the
point of maximal unipotent monodromy given in (\ref{csvariables})
is invariant under
$a_0\rightarrow - a_0$ and the result above can be used {\it verbatim} as
describing a subfamily of the Enriques surface. Let us first
calculate in the Enriques fibre direction
from (\ref{f1top}), which we parameterize as
\be
F^{(1),E}(q_1,q_2)=
r_1\log\det\left(\partial  z_i\over \partial t_j\right)
+ r_2 \log(\hat X^0) + r_3\log(z_1 z_2) + r_4\log \Delta.
\label{F1enriques}
\ee
Notice that this can be also interpreted as a
calculation for K3$\times \bT^2$ in which one would expect
that $F^{(1)}$ vanishes. This is the case if we set
$r_1= -r_2$, $r_2=r_3=r_4$.
As we will explain in subsection 6.3.3, the heterotic prediction (\ref{heteroticprediction})
is reproduced for the choice
$r_1=-{1\over 2}(2+r_2)$, $r_3={1\over 2} (2+r_2)$ and
$r_4={1\over 4}(1+r_2)$. Comparing with (\ref{f1top}) we set $r_2=-3$ to get $r_1={1\over 2}$.
This yields also the expected result
$\kappa=3+h_{11}-{\chi\over 12}=6$ for the three parameter
model.

We can now use the arithmetic expressions for the mirror map (\ref{analyticmirrormap})
and fundamental period (\ref{anxo}) to derive an exact analytic B-model expression
for the Ray-Singer torsion $F^{(1)E}(q_1,q_2)=F^{(1)}(q_1,q_2, q_3=0)$ in the Enriques direction
(\ref{F1enriques})
\be
F^{(1)\, E}=-{1\over 2} \log(\delta/16),
\ee
where
\be
\delta=K_2^2 \tilde K_2^2 - K_4 \tilde K_2^2 - K_2^2\tilde K_4
\ee
and we used the fact that the discriminant (\ref{discriminant}) can be written as
\be
\label{disceasy}
\Delta=\left(\delta \over K_2^2 \tilde K_2^2\right)^2.
\ee
Indeed, we can write as well
\be
\label{fonedisc}
F^{(1)\, E}=-{1\over 4} \log (\Delta) -2 \log (\hat X^0).
\ee
On the B-model side it is easy to argue that, in genus one,
there are no contribution
from curves in classes with mixed degree in base and fiber.
According to (\ref{mirrormapK3},\ref{mirrormapT2}) the term
$\det \left(\partial t / \partial z\right)$ will factorize. The
same is true for the $X^0$ contribution due to (\ref{X0}). Finally
the $f_1(z)$ is a product of discriminant factors. There will be two for the
fibre and one for the base with no mixing between the base and fibre
complex structure coordinates. Because of the logarithm we get a
sum of two terms. The first one is the $F^{(1) E}$ in the fiber, and depends
only on $q_i$, $i=1,2$ while the second one depends only of the $\tau$ parameter
of the base, $q_3=\exp(2 \pi i \tau)$. We conclude that
the total Ray-Singer torsion for the FHSV model is
\be
\label{gonefhsv}
F^{(1)}(q_1,q_2,q_3)=-{1\over 2} \log(\delta/16) -12 \log(\eta(q_3)),
\ee
where the contribution of the base is the same one as for K3$\times \bT^2$. This has
also been argued in \cite{hmborcherds}, however the expansion in \cite{hmborcherds}
is not related to the instanton expansion of the type IIA string as explained in sections
\ref{sec:geometricreduction} and \ref{sec:borcherds}.

\subsubsection{Propagators and higher genus amplitudes}

In order to compute the $F^{(g)}$ amplitudes for $g>1$ we use the holomorphic anomaly equations of
\cite{bcov}. These equations lead to an expression for $F^{(g)}$ of the form
\be
\label{generalst}
F^{(g)}=F^{(g)}_{\rm rec} + (X^0)^{2g-2} f_g,
\ee
where $F^{(g)}_{\rm rec}$ is completely determined in a recursive way from the topological string amplitudes
at lower genera $F^{(g')}$, $g'<g$, their derivatives w.r.t. the flat coordinates, and the propagators of the Kodaira-Spencer
theory introduced in \cite{bcov}. $f_g$ is the holomorphic ambiguity at genus $g$. It is a rational function on the moduli space
of complex structures and to determine it we need some extra data, like for example explicit values of Gromov-Witten invariants at low
degree. 

If we apply this procedure to the reduced, three--parameter model, we find important simplifications in the computation of $F^{(g)}_{\rm rec}$. This is due to the fact that,
in flat coordinates, there is only one nonzero Yukawa coupling $C_{123}$ which moreover does not receive any worldsheet instanton corrections
and it is simply given by $C_{123}=1$. Further derivatives of the Yukawa coupling vanish, and this sets to zero many
contributions to $F^{(g)}_{\rm rec}$.

One of the fundamental ingredients of the holomorphic anomaly equations of \cite{bcov} are the
propagators $S^{ij}$, $S^i$ and $S$ of Kodaira-Spencer theory, where $i,j$ are indices for the complex moduli. The procedure to find these propagators in the
multiparameter case
has been explained in \cite{kkrs}. It turns out that, in the case of the reduced model that we are studying, one can make a choice of gauge in which
they have a particularly simple form. One first finds that it is possible to set $S^{ii}=0$, $i=1,2,3$. To solve
for the remaining propagators with two indices $S^{ij}$, $i\not= j$, one can use the equation
\be
\label{propader}
{1\over 2}S^{ij} C_{ijk} = \partial_k F^{(1)} + \biggl( {\chi\over 24} -1 \biggr)\partial_k ( \log X^0 + \log f),
\ee
where $f$ is a holomorphic function of the complex moduli which arises as an integration constant of the
holomorphic anomaly equations. The equation (\ref{propader}) determines uniquely
the propagators $S^{12}$, $S^{13}$, $S^{23}$ up
to an integration constant. Using now (\ref{fonedisc}), we see that the only piece of $S^{13}$ that
cannot be absorbed into $f$ is $-\partial \log  \hat X^0/\partial t_2$. The same thing happens to $S^{23}$, after
exchanging $t_1 \leftrightarrow t_2$. We then make the following
choice of propagators
\be
\ba
S^{13}=&- {1\over 2} {d \over dt_2} \log\,\tilde K_2 = {1\over 2} \partial_{t_2} F^{(1)} +   A_2, \\
S^{23}=&- {1\over 2} {d \over dt_1} \log\,K_2 = {1\over 2} \partial_{t_1} F^{(1)} +  A_1,
\ea
\ee
where
\be
\label{ais}
A_i ={1\over 8}\partial_{t_i} \log \, \Delta, \qquad i=1,2.
\ee
Finally, one can choose the remaining propagator $S^{12}$ to be $-E_2(q_3)/12$.
The final result for the propagators $S^{ij}$ of the reduced model is then
\be
\label{propa}
\begin{aligned}
S^{13}=&S^{31}={1\over 12} E_2(q_2) +{1\over 8}{\tilde K_4\over \tilde K_2} -{1\over 24} \tilde K_2 ,\\
S^{23}=&S^{32}=S^{13}(q_1\leftrightarrow q_2),\\
S^{12}=&S^{21}=-{1\over 12}E_2 (q_3),\\
S^{11}=&S^{22}=S^{33}=0.
\end{aligned}
\ee
Notice that, with this choice, $S^{ij}$ only depends on $q_k$, where $(ijk)$ is a permutation of $123$.
By using now the explicit expressions in \cite{bcov}, it is easy to see that the propagators $S^i$, $S$ can be
chosen to be
\be
S^i =S^{ij}S^{ik},  \quad i=1,2,3, \qquad S=S^{12} S^{13}S^{23},
\ee
where $(ijk)$ is a permutation of $123$. Notice that the structure of the propagators of the
reduced model is very similar to the case of toroidal orbifolds studied in \cite{bcov}.

Using the one-loop result (\ref{gonefhsv}), the recursive formula of \cite{bcov} for $F^{(2)}$
leads to the expression
\be
\label{f2fiberB}
\begin{aligned}
F^{(2)}=& S^{12}(F^{(1)}_{12} + F^{(1)}_{1} F^{(1)}_{2}) -2 S^{12} (S^{13}F^{(1)}_{1}+ S^{23} F^{(1)}_{2}) +
4 S^{12} S^{13}S^{23} \\
& -{1\over 2} E_2(q_3) (S^{13}F^{(1)}_{1}+ S^{23} F^{(1)}_{2} - 2 S^{13} S^{23})+ (X^0)^2 f_2,
\end{aligned}
\ee
where $F^{(1)}_i$ denote derivatives of $F^{(1)}$ w.r.t. $t_i$, $X^0=\hat X^0 (q_1,q_2) x^0 (q_3)$,
and $f_2$ is the holomorphic ambiguity. In the fibre limit
\be
S^{12} \rightarrow -{1\over 12}, \quad X^0 \rightarrow \hat X^0, \quad f_2 \rightarrow f_2^E,
\ee
(\ref{f2fiberB}) should become
$F^{(2)E}$, the genus two amplitude on the fiber. Here $f_2^E$ is simply the fibre limit of the
holomorphic ambiguity. The heterotic prediction can be
recovered by simply setting
\be
f^E_2=0.
\ee
After using the explicit expressions for the propagators (\ref{propa}) and taking into account that
\be
F^{(1)}_{12} = 8 A_1 A_2,
\ee
where the $A_i$ are defined in (\ref{ais}), we find that
the genus two free energy for the Enriques fibre is simply given by
\be
\label{twofiber}
F^{(2)\, E}=-{1\over 4} F^{(1)}_1 F^{(1)}_2,
\ee
where $F^{(1)}_i$ can be written in terms of
$K_i$, $\tilde K_i$, $E_2=E_2(q_1)$ and $\tilde E_2=E_2(q_2)$, as
\be
F^{(1)}_i={1\over 6} \Bigl(E_2 (q_i)-{\kappa K_2 (q_i) \over 2 \delta } \Bigr),
\quad i=1,2,
\label{f1delta}
\ee
and $\kappa={\delta+3 K_4 \tilde K_4}$. 

In the computation of $F^{(2)}$ we have used the formulae for $F^{(g)}_{\rm rec}$ obtained in \cite{bcov}, and we have 
applied them directly to the reduced model. However, for higher genus amplitudes this method is problematic. The reason for this is that, 
in order to obtain the correct result, we have to implement the reduction consistently, i.e. we have to first consider the formulae for $F^{(g)}_{\rm rec}$ in the 
original model with $11$ K\"ahler parameters and then set the $E_8(-1)$ parameters to zero. In general, the result of this will be different from the result obtained by 
computing $F^{(g)}_{\rm rec}$ directly in the reduced, three--parameter model. The two procedures lead to the 
same tensorial structures, but with different numerical coefficients. It can be seen that at genus $2$ this is not a problem, 
but for higher genus we cannot use the reduced model to obtain the answer for $F^{(g)}$.

One can still use the recursive formulae in the reduced model in order to obtain general properties of the amplitudes, as well 
as expressions for $F^{(g) E}$ in terms of modular forms. For 
$g=3$ one finds, for example,  
\be
\begin{aligned}
F^{(3)\, E}=&-{1\over 2^{10} 3^4}\Biggl( \left(E_2^2-{\kappa E_2 K_2 \over \delta}  +{\rho_1\over (2\delta)^2} \right)
\left(\tilde E_2^2 -{\kappa \tilde E_2 \tilde K_2 \over \delta}   +{\rho_2 \over (2 \delta)^2}\right) \\
&\ds{+ 9 \mu \left({1\over \delta^2}(E_2 \tilde K_2 -\tilde E_2  K_2)^2 - {3\over 4 \delta^4}\rho_3 \right) \Biggr)},
\end{aligned}
\ee
where $\mu=(K_2^2-K_4)(\tilde K_2^2-\tilde K_4) K_4 \tilde K_4$ and
\be
\begin{aligned}
\rho_1&=\kappa^2  K^2_2 -
9 ( K_2^2 - K_4) K_4 ((\tilde K_2^2 + 7 \tilde K_4)\delta +9 \tilde K_2^2 K_4 \tilde K_4),\\
\rho_2&=\rho_1(q_1\leftrightarrow q_2),\\
\rho_3&=\delta^3 + (7 \delta^2 + 72 \delta K_4 \tilde K_4)(K_2^2 \tilde K_4+ K_4 \tilde K_2^2)
+12 K_4 \tilde K_4 (5 \delta^2+6 K_2^2 \tilde K_2^2 K_4 \tilde K_4).
\end{aligned}
\ee
We have also found explicit results for the genus four topological string amplitude, which are too long to be
reported here but are available on request. Note that $F^{(g)E}$ exhibits a leading order pole at the discriminant
$\Delta$ of the form
\be
F^{(g)E}\sim {b_g\over \Delta^{g-1}}={\tilde b_g\over \delta^{2g-2}} \ ,
\ee
which restricts the ansatz for the holomorphic ambiguity along the fiber $f^E_g$.

\subsubsection{Comparison with the heterotic results}
\begin{table}
\begin{center}
\begin{tabular}{| l | r  r r  r r |}
\hline
$m$&$n=0$&  1          &2      &3       &4  \\ \hline
0&-& 8& 4& 8& 4  \\
1&8& 2048& 49152& 614400& 5373952  \\
2&4& 49152& 5372928& 216072192& 5061451776  \\
3& 8& 614400& 216072192& 21301241856& 1063005978624  \\
4&4& 5373952& 5061451776& 1063005978624& 100372720320512  \\
5&8& 37122048& 83300614144& 34065932304384& 5641848336678912  \\
6&4& 216072192&1063005671424& 794110053826560& 218578429867425792 \\
\hline
\end{tabular}
\caption{Genus one  BPS invariants $n_1(m,n)$ for the two parameter subfamily. By construction
there is a symmetry under exchange of $(m,n)$.}
\end{center}
\label{genusonetab}
\end{table}
Let us now do a more detailed comparison of the B-model results with the heterotic predictions in the
geometric reduction. Table 7 contains the B-model
prediction for genus one BPS numbers in
the classes $(m,n)$ of $\Gamma^{1,1}$, the
cohomology lattice of the reduced model. This is identified with the $\Gamma^{1,1}$ sublattice inside
the Picard lattice of the Enriques surface $\Gamma_E=\Gamma^{1,1}\oplus E_8(-1)$.
The total class in $\Gamma^{1,1}\oplus E_8(-1)$ is labelled by a vector $r=(n,m,\vec v)$ with norm $r^2=2 m n -  \vec v^{~2}$.
The B-model for the subfamily calculates BPS invariants
in which, for given $(m,n)$, one sums over all possible vectors $\vec v$
in the $E_8(-1)$ lattice. Recall that the coefficients of $q$ in the $E_8$-theta function
\begin{equation}
\label{thetae8}
\Theta_{E_8}(q)=\sum_{ \vec v\in E_8(1)}
q^{\vec v ^{2}/2}=\sum_{\vec v^{2}} m(\vec v^{2}/2) q^{\vec v^{2}/2} =E_4(q)=1+240 q+2160 q^2+ 6270 q^3+\ldots
\end{equation}
yield the total number $m(\vec v^{2}/2)$ of vectors $\vec v$ with a given norm. This means that the results obtained from the reduced B model
should correspond to a ``reduced'' heterotic theory in which we freeze the moduli
of the $E_8(-1)$ lattice. Therefore, the topological string amplitudes of this ``reduced" heterotic model will be given by
\be
F^{\rm red}_g(t)=\sum_{n,m} c^{\rm red}_g(2 n m ) \bigg\{ 2^{3-2g} {\rm Li}_{3-2g}(q_1^n q_2^m) -
{\rm Li}_{3-2g}(q_1^{2n} q_2^{2m})\biggr\},
\label{redprediction}
\ee
where
\be
\label{geomredmod}
\sum_n c^{\rm red}_g(n) q^n =f_1(q)E_4(2\tau) {\cal P}_{g}(q).
\ee

The first thing we notice when we compare the heterotic theory and the B-model is that, in the results from the B model, the
BPS invariants depend only on the product $n m$ but also on whether the
class $(n,m)$ is even or odd. For $(m,n)$ in $\Gamma^{1,1}$ to be in a even class,
$m$ and $n$ have to be even. In odd classes either $m$ or $n$ or both
are odd. This is needed in order to match (\ref{redprediction}). One can easily see that indeed
there is a precise agreement between (\ref{redprediction}) and the B-model results presented above. For example,
the invariant $2048$ at genus one in the B-model is given by
\be
2048=128 + 240\cdot 8,
\ee
where the number $240$ counts the vectors of norm $\vec v^{2}=2$. Notice that, as a by-product of this
comparison at $g=1$, we obtain the following Borcherds-type identity
\be
K_2^2 (q_1)K_2^2(q_2) - K_4(q_1) K_2^2(q_2) - K_2^2(q_1)K_4 (q_2) =
 16 \prod_{n,m} \biggl( { 1-q_1^n q_2^m \over
1+q_1^{n} q_2^{m}}\biggl)^{c(2 nm)},
\ee
where
\be
\sum_n c(n) q^n =-{64 \over 3 \eta^6(\tau) \vartheta_2^6(\tau)} E_2(\tau)E_4(2\tau).
\ee

One can also check that the above B-model expressions on the fiber, for $2\le g\le 4$, agree with the heterotic prediction for the reduced model
(\ref{redprediction}).

\subsubsection{Extending the results to the CY threefold}

Let us now turn from the fibre limit to the full Enriques Calabi-Yau by including
the base classes. We first notice one important general property of $F^{(g)}$: it will be a modular form with respect
to a modular subgroup\footnote{It was noticed in \cite{Yamaguchi:2004bt} that the $F^{(g)}$ 
calculated for the quintic in $\IP^4$ in \cite{kkv}  can be written as polynomials of five generators. 
This is presumably a manifestation of modular properties of the correponding $F^{(g)}$ w.r.t. 
the modular group of the quintic in $SP(4,\ZZ)$.} in ${\rm SL}(2,\ZZ) 
\times {\rm SL}(2,\ZZ) \times  {\rm SL}(2,\ZZ)$, acting on the
parameters $(q_1, q_2, q_3)$. The modular weight is given by
\be
\label{modu}
(2g-2,2g-2,2g-2).
\ee
This can be seen most easily by looking at the last term in (\ref{generalst}). The holomorphic ambiguity has zero modular
weight, since it is given by a rational function of the coordinates $z_1, z_2, z$, which are all modular forms of zero weight.
The fundamental period $X^0$ has however modular weight $(1,1,1)$, which leads to (\ref{modu}).

In order to study the full dependence on the $\bT^2$ on the base, we first notice that the
$F^{(g)}$ amplitudes restricted to the base vanish, since they should be equal to the topological
string amplitudes of K3$\times \bT^2$ on the base. If we denote $F^{(g)B}(q_3) =F^{(g)}(q_1=0,q_2=0,q_3)$, we have,
\be
\label{basetwo}
F^{(g)B}=0, \qquad g>1.
\ee
We now use this information to find expressions for $F^{(g)}(q_1,q_2,q_3)$. We first consider $g=2$.
Since $S^{13}={\cal O}(q_2)$, $S^{23}={\cal O}(q_1)$, the expression (\ref{f2fiberB}) reproduces
the result (\ref{basetwo}) if we assume that the holomorphic ambiguity vanishes as well when restricted to the base.
Given these facts, it is natural to assume that the total ambiguity $f_2$ vanishes for the choice of
propagators (\ref{propa}). In this case, the only dependence of (\ref{f2fiberB}) on $q_3$ is an overall factor
$E_2(q_3)$, and we finally obtain the remarkably simple expression for $g=2$,
\be
\label{finalf2}
F^{(2)}=E_2(q_3) F^{(2)E}.
\ee
This simple form for the reduced model agrees with the results of Maulik and Pandharipande on the Gromov-Witten
invariants at genus two for mixed classes of the full Enriques CY \cite{mp} (in fact, their results
suggested to us the existence of a simple gauge for the propagators). Notice that (\ref{finalf2}) can be also written, after setting 
$q_3=\re^{-t_3}$, as
\be
 \label{finalf2bis}
F^{(2)}=-{1\over 2} F_1^{(1)}F_2^{(1)}F_3^{(1)},
\ee
again with the same structure as the genus 2 amplitude for the toroidal orbifold considered in \cite{bcov}.

As is clear from (\ref{f1delta}), the expression (\ref{finalf2bis}) exhibits a pole
at ${1\over \Delta}\sim {1\over \delta^2}$. A remarkable fact is that it does not
have such a pole at the discriminant $\Delta_b=1-432 z_3$ of the base curve (\ref{basecurve}).
It seems reasonable to assume that such poles do not occur at any genus and to refine (\ref{modu})
in that $F^{(g)}$ is a {\sl holomorphic}, quasimodular function
of $q_3$ of weight $2g-2$, i.e. in what concerns the dependence on $q_3$
it is generated by $E_2,E_4,E_6$.  In this respect, the modular properties of $F^{(g)}$ with respect to
the modular parameter of the torus would be similar to those found in the case of
the elliptic curve \cite{DijkgraafZagier}. On the other hand, $F^{(g)}$ is a {\it weakly holomorphic}
function of $q_1,q_2$ with weights $(2g-2,2g-2)$, which in particular contains $\delta^{-1}$. 
This assumption restricts the ambiguity considerably and leads uniquely
to the following expression for $F^{(3)}$:
\be
F^{(3)}=E^2_2(q_3) F^{(3)E}  + (E_2^2(q_3) - E_4 (q_3)) H^{(3)} (q_1, q_2),
\label{totalfthree}
\ee
where
\be
\label{h3}
H^{(3)}(q_1, q_2) = -{1\over 2} F^{(3)E} -{1\over 24} (F_1^{(1)E} F_2^{(2)E} +F_2^{(1)E} F_1^{(2)E} ).
\ee
In the case $g=3$ we don't have results on the mixed classes to compare with and check in detail the
conjectures (\ref{totalfthree}) and (\ref{h3}). However, we have verified
that they lead to results which are consistent with (\ref{basetwo})
and with integrality of the BPS
numbers $n_g(r) \in \ZZ$ in the expansions (\ref{gvexpansion}). As expected
from the discussion above (\ref{eq:adjunction}), all of these numbers
are divisible by eight. In the tables above 
we list some BPS invariants $n_g(r)$ for base degree $d$ equal to one. 
In \cite{gkmw}, we will extend these ideas and present an alternative and more 
powerful method to derive (\ref{totalfthree}) and (\ref{h3}) 
which makes also possible to obtain results for $F^{(g)}$ to high genus.

\begin{table}
\begin{center}
\begin{tabular}{| l | r  r r  r r |}
\hline
$m$&$n=0$&  1          &2      &3       &4  \\ \hline
0&0& 0& 0& 0& 0  \\
1&0&384&99072&2557440& 34604544 \\
2&0&99072&34604544&2425752576&82015423488\\
3&0&2557440&2425752576&399200753664&28156719273984    \\
4&0&34604544&82015423488&28156719273984&3717898174470144 \\
\hline
\end{tabular}
\caption{Genus two BPS invariants $n_2 (m,n,1)$ for branes wrapping
the base torus once.}
\end{center}
\label{genus2wrap1}
\begin{center}
\begin{tabular}{| l | r  r r  r r |}
\hline
$m$&$n=0$&  1          &2      &3       &4  \\ \hline
0&0& 0& 0& 0& 0  \\
1&0&128&33792&10521600&17047552 \\
2&0&33792&25704448&2596196352&113305067520\\
3&0& 1052160&2596196352&635491780608&58963231506432  \\
4&0&17047552&113305067520& 58963231506432&10321183934611456  \\
\hline
\end{tabular}
\caption{Genus three BPS invariants $ n_3 (m,n,1)$ for branes wrapping
the base torus once.}
\end{center}
\label{genus3wrap1}
\end{table}

\section*{Acknowledgments}
We would like to thank Ignatios Antoniadis, Jim Bryan, Igor Dolgachev, Rajesh Gopakumar,
Sheldon Katz, Wolfgang Lerche, Margarida Mendes Lopes, Boris Pioline,
Emanuel Scheidegger, and Cumrun Vafa for useful
conversations, and Ciprian Borcea and Elias Kiritsis for useful correspondence.
We are specially grateful to Greg Moore for extensive discussions on various
issues addressed in this paper, and for a detailed critical reading of the manuscript. Finally, our thanks to
Davesh Maulik and Rahul Pandharipande for extensive discussions
on the Gromov--Witten theory of the Enriques CY, and for sharing their results with us.

\appendix
\sectiono{Theta functions and modular forms}
Our conventions for the Jacobi theta functions are:
\be
\begin{aligned}
\vartheta_1(\nu|\tau)&=\vartheta [^1_1](\nu|\tau)=
i \sum_{n \in {\bf Z}} (-1)^n q^{{1\over 2}(n+1/2)^2} e^{i \pi (2n +1) \nu},\\
\vartheta_2(\nu|\tau)&= \vartheta [^1_0](\nu|\tau)=
\sum_{n \in {\bf Z}} q^{{1\over 2}(n+1/2)^2} e^{i \pi (2n +1) \nu},\\
\vartheta_3(\nu|\tau)&= \vartheta [^0_0](\nu|\tau)=
\sum_{n \in {\bf Z}} q^{{1\over 2} n^2} e^{i \pi 2n  \nu},\\
\vartheta_4(\nu|\tau)&= \vartheta [^0_1](\nu|\tau) =
\sum_{n \in {\bf Z}} (-1)^n q^{{1\over 2}n^2} e^{i \pi 2n  \nu},
\end{aligned}
\ee
where $q=e^{2\pi i \tau}$. When $\nu=0$ we will simply denote $\vartheta_2(\tau)=
\vartheta_2(0|\tau)$ (notice that $\vartheta_1(0|\tau)=0$).
The theta functions $\vartheta_2(\tau)$, $\vartheta_3(\tau)$ and $\vartheta_4(\tau)$ have the
following product representation:
\be
\begin{aligned}
\vartheta_2(\tau)&  =2 q^{1/8}\prod_{n=1}^{\infty} (1-q^n)(1+q^n)^2,\\
\vartheta_3(\tau)&  = \prod_{n=1}^{\infty} (1-q^n)(1+q^{n-\half} )^2,\\
\vartheta_4(\tau)& = \prod_{n=1}^{\infty} (1-q^n)(1-q^{n-\half} )^2
\end{aligned}
\ee
and under modular transformations they behave as:
\be
\begin{aligned}
\vartheta_2 (-1/\tau)= &{\sqrt { \tau \over i}} \vartheta_4 (\tau),\\
\vartheta_3 (-1/\tau)= &{\sqrt { \tau \over i}} \vartheta_3 (\tau),\\
\vartheta_4 (-1/\tau)= &{\sqrt { \tau \over i}} \vartheta_2 (\tau),
\end{aligned}
\quad\quad
\begin{aligned}
\vartheta_2 (\tau +1)= &{\rm e}^{i\pi/4} \vartheta_2 (\tau),\\
\vartheta_3 (\tau +1)= &\vartheta_4 (\tau),\\
\vartheta_4 (\tau +1)= & \vartheta_3 (\tau).
\end{aligned}
\ee
The theta function $\vartheta_1(\nu|\tau)$ has the product representation
\be
\label{prodone}
\vartheta_1(\nu|\tau)=-2 q^{1\over 8} \sin (\pi \nu) \prod_{n=1}^{\infty} (1-q^n) (1-2 \cos (2 \pi \nu) q^n + q^{2n}).
\ee
We also have the following useful identities:
\be
\label{sumthet}
\vartheta_3^4 (\tau) = \vartheta_2^4 (\tau) + \vartheta_4^4 (\tau),
\ee
and
\be
\label{prodtheta}
\vartheta_2 (\tau)\vartheta_3 (\tau)\vartheta_4 (\tau) = 2\, \eta^{3}(\tau),
\ee
where
\be
\label{dede}
\eta(\tau)= q^{1/24} \prod_{n=1}^{\infty} (1- q^n)
\ee
is the Dedekind eta function. One has the following doubling formulae,
\be
\label{deta}
\begin{aligned}
\eta(2\tau)=&{\sqrt {\eta(\tau)\vartheta_2(\tau) \over 2}}, \qquad
\vartheta_2(2 \tau) ={\sqrt { {\vartheta_3^2(\tau) -\vartheta_4^2(\tau) \over 2}}}, \\
\vartheta_3(2 \tau) =&{\sqrt { {\vartheta_3^2(\tau) +\vartheta_4^2(\tau) \over 2}}},\qquad
\vartheta_4(2 \tau) ={\sqrt { \vartheta_3(\tau) \vartheta_4(\tau)}},\\
\eta(\tau/2)=&{\sqrt {\eta(\tau) \vartheta_4(\tau)}}.
\end{aligned}
\ee
The Eisenstein series are defined by
\be
\label{geneis}
E_{2n}(q)=1-{4n \over B_{2n}}\sum_{k=1}^{\infty}{k^{2n-1} q^{k}\over 1-q^{k}},
\ee
where $B_{m}$ are the Bernoulli numbers.
The formulae for the derivatives of the theta functions are also useful:
\be
\begin{aligned}
q {d \over dq}\log\, \vartheta_4=&{1\over 24}\biggl( E_2 -\vartheta_2^4 -\vartheta_3^4\biggr),\\
q {d \over dq}\log\, \vartheta_3=&{1\over 24}\biggl( E_2 +\vartheta_2^4 -\vartheta_3^4\biggr),\\
q {d \over dq}\log\, \vartheta_2=&{1\over 24}\biggl( E_2 +\vartheta_3^4 +\vartheta_4^4\biggr),
\end{aligned}
\ee
and from these one finds,
\be
q {d \over dq}\log\, \eta ={1\over 24}E_2(\tau),\qquad q {d \over dq}E_2={1\over 12}\Bigl( E_2^2 - E_4\Bigr).
\ee
The doubling formulae for $E_2(\tau), E_4(\tau)$ are
\be
\label{deis}
\begin{aligned}
E_2(2 \tau)= &{1\over 2} E_2(\tau) + {1\over 4}(\vartheta_3^4(\tau) +\vartheta_4^4(\tau)),\\
E_4(2 \tau)= & {1\over 16} E_4(\tau) + {15\over 16}\vartheta_3^4 (\tau)\vartheta_4^4 (\tau).
\end{aligned}
\ee

\sectiono{Lattice reduction}

In this Appendix we briefly review the computation of integrals of the form (\ref{Fgint})
with the technique of lattice reduction. This technique applies to integrals of the form
\be
\label{thetatrans}
\int_{\cal F} {d^2 \tau \over \tau_2^2} \sum_J f_J(\tau,\bar \tau) 
{\overline \Theta}_{\Gamma_J}(\tau, \alpha_J, \beta_J, P,\phi)
\ee
These integrals are sometimes called {\it theta transforms} of the (quasi)modular forms $f_J(\tau,\bar \tau)$.
The generalized Siegel-Narain theta function which appears in this integral is defined as
\be
\label{sntheta}
\begin{aligned}
&\Theta_{\Gamma} ( \tau,\alpha, \beta,P,\phi )= \\
&\sum_{p \in \Gamma} \exp \biggl[ -{\Delta
\over 8 \pi \tau_2 }
\biggl] \bigl( \phi(P(\lambda)) \bigr) \exp \biggl[ \pi i \tau (p+\beta/2)_+^2 +
\pi i{\overline \tau}  (p+\beta/2)_-^2 +
\pi i (p+\beta/2, \alpha)\biggr],
\end{aligned}
\ee
where $\Gamma$ is a lattice of signature $(b^+, b^-)$, $P$ is the projection,
$\phi$ is a polynomial on ${\IR}^{b^+,b^-}$
of degree $m^+$ in the
first $b^+$ variables and
of degree $m^-$ in the second $b^-$ variables, and $\Delta$ is the
(Euclidean) Laplacian in ${\IR}^{b^++b^-}$. The rest of the notations where introduced in section 2.
The lattices involved in (\ref{thetatrans}) have all
the same signature, and they only differ in overall factors for their norms as well as in the shifts
$\alpha_J, \beta_J$. In the computation of these integrals
by lattice reduction, one proceeds iteratively and in each step the rank of the lattice is reduced by two.
Proceeding in this way, one can reduce the computation of (\ref{thetatrans}) to evaluation of quantities
associated to the reduced lattices.

Let us consider the simple case in which there is only one term in the sum (\ref{thetatrans}), with $\alpha=\beta=0$,
and the lattice $\Gamma$ is even and self-dual. In this case we will denote (\ref{sntheta}) by
$\Theta_{\Gamma}(\tau, P,\phi)$. The theta transform is then given by
\be
\label{simpletrans}
\Phi_{\Gamma} (P, \phi, F^{\Gamma}) = \int_{\cal F} {d^2 \tau \over \tau_2^2}
{\overline \Theta}_{\Gamma} (\tau,P,\phi) F^{\Gamma} (\tau, \bar \tau),
\ee
where
\be
\label{modform}
F^{\Gamma} (\tau, \bar \tau)= \tau_2^{b^+/2+m^+}F(\tau)
\ee
is a (quasi)modular form with weight $(-b^-/2-m^-, -b^+/2-m^+)$, constructed from the (quasi)modular form
$F(\tau)$ with weights $(b^+/2+m^+-b^-/2-m^-,0)$. We will assume that $F(\tau)$
is an almost holomorphic form, {\it i.e.} it has the expansion
\be
\label{expan}
F(\tau) = \sum_{m\in {\bf Q}} \sum_{t\ge 0} c(m,t) q^m \tau_2^{-t}
\ee
where $c(m,t)$ are complex numbers which are zero for all but a finite number
of values of $t$ and for sufficiently small values of $m$.
Lattice reduction is then implemented as follows. Let $z$ be a
primitive vector of $\Gamma$ of  zero norm, and let $K=(\Gamma \cap z^{\bot})/ {\ZZ} z$. This lattice, which
has signature
$(b^+ - 1, b^--1)$, is called the reduced lattice. A typical situation when choosing a reduction vector occurs when the
lattice $\Gamma$ has $\Gamma^{1,1}$ as a sublattice, where $\Gamma^{1,1}$ is the lattice of rank two and intersection form
\be
\left( \begin{matrix} 0 & 1 \\
                             1  & 0  \end{matrix} \right).
\ee
In this case, one can take $z$ to be one of the vectors that generate $\Gamma^{1,1}$.
In the reduced lattice one can construct ``reduced" projections $\widetilde P$ as
follows: consider $z_{\pm} \equiv P_{\pm} (z)$, and decompose ${\IR}^{b^{\pm}}
\simeq \langle z_{\pm}  \rangle \oplus \langle z_{\pm} \rangle ^{\bot}$. The
projection on the orthogonal complement $\langle z_{\pm} \rangle ^{\bot}$
is the reduced projection $\widetilde P_{\pm}$. It can be explicitly written in
terms of $P_{\pm}$ as
\be
\widetilde P_{\pm} (p) = P_{\pm} (p) - { ( P_{\pm} (p),
z_{\pm}) \over
z^2_{\pm}}  z_{\pm}.
\ee
Once this reduced projection has been constructed, we have to decompose the
polynomial
involved in (\ref{sntheta}) with respect to this projection, according to the
expansion
\be
\label{poldec}
\phi(P(p))= \sum_{h^+, h^-} (p, z_+)^{h^+} (p, z_-)^{h^-}
\phi_{h^+, h^-} (\widetilde P (p)),
\ee
where $p_{h^+, h^-}$ are homogeneous polynomials of degrees $(m^+-h^+,
m^--h^-)$ on $\widetilde P (\Gamma \otimes \IR )$. We now write the vectors
of the lattice $\Gamma$ as
\be
p=c z' + mz + p^K,
\ee
where $p^K$ is a vector in the reduced lattice $K$ and $(z',z)=N$.
When the reduction vector belongs to a sublattice $\Gamma^{1,1}$, the
vector $z'$ is the other generator of the sublattice and $N=1$. One can now rewrite the Siegel-Narain
theta function in terms of the reduced lattice, after a Poisson resummation of the integer $m$, as
\cite{borcherdstwo}
\be
\label{thetared}
\begin{aligned}
& \Theta_{\Gamma}(\tau,P,\phi)=
{1 \over \sqrt {2 \tau_2 z_+^2} } \sum_{h \ge 0} \sum_{h^+, h^-} {h!
\over (-2i\tau_2)^{h^+
+h^-}} \biggl( -{\tau_2 z_+^2
\over \pi } \biggr)^h {h^+ \choose h}{h^- \choose h} \\
&\times \sum_{c,\ell} (N c \bar \tau + \ell)^{h^+-h}( N c  \tau + \ell)^{h^--h}
\exp\biggl(-{\pi |N c \tau + \ell|^2 \over 2 \tau_2 z_+^2}\biggr) \Theta_{K}(\tau,  \ell \mu/N, -c\mu,  \widetilde P,
\phi_{h^+, h^-}).
\end{aligned}
\ee
In writing this formula we have assumed that $(p_K,z')=(z',z')=0$, and we have introduced the vector
\be
\mu= -z' + N \biggl({z_+ \over 2 z_+^2} + { z_- \over 2 z_-^2}\biggr) \in K\otimes \IR.
\ee
Once this expression is inserted into the integral (\ref{thetatrans}), one can apply the
``unfolding'' procedure, in which the integral over the fundamental domain ${\cal F}$ of ${\rm SL}(2,\ZZ)$ becomes an integral over
the domain $[-1/2,1/2]\times (0,\infty)$. At the same time, one can set $c=0$ in (\ref{thetared}) by modular invariance.
There are two types of contributions in the end. The first one comes from $\ell=0$ and it is sometimes called the contribution of the
``zero orbit.'' It is given by
\be
\label{degbor}
{1 \over {\sqrt {2 z_+^2}} } \sum_{h\ge 0} \biggl( {z_ +^2 \over 4\pi }\biggr)
^h \Phi_K({\widetilde P} ,
\phi_{h,h} , F^K).
\ee
Notice that this is another theta transform, but for the reduced lattice, which is smaller.
The contribution from the nonzero orbits comes from $\ell>0$, and it involves a sum over the reduced lattice $K$.
When $\widetilde P_+
(\lambda^K)\not=0$, it is given by
\be
\label{nondeg}
\begin{aligned}
&  {\sqrt {2 \over z_+^2} } \sum_{h \ge 0} \sum_{h^+, h^-} {h! \over (2i)^{h^+
+h^-}} \biggl( -{z_+^2
\over \pi } \biggr)^h {h^+ \choose h}{h^- \choose h} \sum_{j} \sum_{p^K
\in K} {1 \over j!} \biggl( -{\Delta \over 8\pi}\biggr)^j \overline p_{h^+,
h^-}(\widetilde P (p^K))  \\
&\cdot  \sum_{\ell=1}^{\infty} \re^{ 2\pi i \ell(p^K, \mu)/N} \sum_t 2 c(p^2/2,t) \biggl( {\ell
\over 2|z_+||\widetilde P_+
(p^K)|} \biggr) ^{h-h^+-h^- -j -t+b^+/2+m^+-3/2}\\
& \cdot  K_{h-h^+-h^--j-t+b^+/2+m^+-3/2}
\biggl( {2\pi \ell |\widetilde P_+
(p^K)|\over  |z_+|} \biggr).
\end{aligned}
\ee
Here, $K_\nu (z)$ is the modified Bessel function, which comes from an integral over
the strip $\tau_2>0$.  When  $\widetilde P_+
(\lambda^K)=0$, the integral has to be regularized and this leads to
a slightly different expression which can be found
in \cite{borcherdstwo}.

An important remark is that the above expressions are only
valid if
\be
|z_+|^2  \ll 1.
\ee
For a fixed primitive
vector $z$, the value of $z_+^2$ depends on the projection
we choose for the lattice, which is in turn determined by the Narain moduli
of the string compactification. Different choices of regions in moduli space will require
different choices of vectors for the lattice reduction, and therefore to different expressions for the resulting
integral.

We have considered here the simplest case in which the theta transform (\ref{thetatrans})
involves a single summand with $\alpha=\beta=0$. More general cases can be
also analyzed with the technique of lattice reduction, see \cite{mmtwo,lm} for examples.

\end{document}

The generic expression for
$F^{(3)}$ is rather complicated (an explicit expression is written down in \cite{kkv}), but in this case one can find important simplifications. The final expression reads:
\be
\label{genusthree}
F^{(3)} = -{1\over 144} E_2^2(q_3) G^{(3)}(q_1, q_2)  + {1\over 288} (E_2^2(q_3) - E_4 (q_3))
H^{(3)}(q_1, q_2) + E_4 (q_3) (\hat X^0)^4 f_3,
\ee
where $G(q_1, q_2) $, $H(q_1, q_2) $ depend only on the K\"ahler parameters on the fiber. They can be written in terms of $A_i$, $i=1,2$, and derivatives of the
genus one amplitude, $F^{(1)}$, as follows:
\be
\ba
G^{(3)}(q_1, q_2)=&106 \,A_1^2 A_2^2 + 40 A_1 A_2 (A_1 F_2^{(1)} + A_2  F_1^{(1)}) +5( A_1  F_{122} ^{(1)} + A_2  F_{112} ^{(1)}) \\ - &68 \,A_1 A_2 F_1^{(1)} F_2^{(1)}
+{13 \over 2}  \Bigl( A_1^2( (F_2^{(1)})^2  + 2 F_{22} ^{(1)})+
A_2^2 ( (F_1^{(1)})^2 + 2 F_{11} ^{(1)})\Bigr)  \\-&  {23\over 8} (F_1^{(1)} F_2^{(1)})^2
-{23\over 4} \Bigl( (F_1^{(1)})^2 F_{22}^{(1)} + (F_2^{(1)})^2 F_{11}^{(1)} \Bigr) -{5\over 2} F_{11}^{(1)} F_{22}^{(1)} \\
+& {1\over 2} \Bigl( F_1^{(1)} F_{122}^{(1)} + F_2^{(1)} F_{112}^{(1)} \Bigr)+
{1\over 2} F^{(1)}_{1122},\\
H^{(3)}(q_1, q_2)=& 108 \,A_1^2 A_2^2 +3  \Bigl( A_1^2( (F_2^{(1)})^2  + 2 F_{22} ^{(1)})+
A_2^2 ( (F_1^{(1)})^2 + 2 F_{11} ^{(1)})\Bigr)\\
-&24 A_1 A_2 F_1^{(1)} F_2^{(1)}  -  {9\over 4} (F_1^{(1)} F_2^{(1)})^2
-{3\over 2} \Bigl( (F_1^{(1)})^2 F_{22}^{(1)} + (F_2^{(1)})^2 F_{11}^{(1)} \Bigr).
\ea
\ee
In (\ref{genusthree}), $f_3$ is the holomorphic ambiguity. In the fiber limit $q_3 = 0$ one finds
\be
F^{(3)E} = -{1\over 144} G^{(3)}(q_1, q_2)  + (\hat X^0)^4 f^E_3,
\ee
and $f_3^E$ can be determined by comparison to the
heterotic results. One finds a relatively simple expression,
\be
\label{threeambigu}
f_3^E={2\over 3} z_1 z_2 \biggl( 13 + {640(z_1 + z_2) -2\over \Delta} + 262144 {z_1 z_2 \over \Delta^2} \biggr).
\ee

The BPS number $n_4 (1,1,1)=80$ seems accessible to the methods of  section \ref{sec:BPStables}.
As some confirmation of the conjecture (\ref{ambconj}), we notice that $n_4(1,1,1)$ would turn out to be
${1328\over 5}$ if we do not remove
the branch cut (\ref{killbranchcut}) by the factor $1-864 \, z_3$ in (\ref{ambconj}).
We also note that, beyond genus five, vanishing theorems for the BPS numbers give
further restrictions on the ambiguity, which keep the model solvable under the
assumption of holomorphicity in $q_3$ even if the generalization of (\ref{ambconj}) for higher genus turns
out to be too strong.

\begin{center}
\begin{tabular}{| l | r  r r  r r |}
\hline
$m$&$n=0$&  1          &2      &3       &4  \\ \hline
0&0& 0& 0& 0& 0  \\
1&0& 80&    -6720&     -791616&    35275008\\
2&0& -6720&-14236416&-1206259968&-41161253376 \\
3&0& -791616&-1206259968&-400202940160&-32439423925248  \\
4&0& 35275008&-41161253376&-32439423925248&-6132421123474432  \\
\hline
\end{tabular}
\caption{Genus four BPS invariants $n_4 (m,n,1)$ for branes wrapping
the base torus once.}
\end{center}
\label{genus4wrap1}